\documentclass[prx,twocolumn,groupedaddress]{revtex4-2}
\usepackage{amsmath,amsfonts,amssymb,graphicx,xcolor}

\newcommand{\ve}[1]{\mathbf{#1}}

\begin{document}

%\preprint{}

%\title{Computing Spatially Dependent Memory Corrections to the Generalized Langevin Equation in the Presence of a Potential of Mean Force}
\title{Generalized Langevin Equation  with a Non-Linear Potential of Mean Force and Non-Linear Memory Friction   From a Hybrid Projection Scheme}


\author{Cihan Ayaz}
\author{Benjamin A Dalton}
\author{Roland R. Netz}
%\thanks{}
\affiliation{Fachbereich Physik, Freie Universit\"at Berlin}


\date{\today}

\begin{abstract}
% insert abstract here: PRX < 500 words
We introduce a hybrid projection scheme that combines linear Mori projection and conditional Zwanzig projection techniques 
 and use it  to derive  a Generalized Langevin Equation (GLE) %\textcolor{red}{\sout{from the Liouville equation}} 
 for a general interacting many-body system.
The resulting GLE includes
i)  explicitly the  potential of mean force (PMF) that  describes the equilibrium 
distribution of the system in the chosen space of reaction coordinates,
ii) a random force term that explicitly depends on  the initial state of the system,
and iii) a memory  friction contribution that splits into two parts:
a  part that is linear in the past reaction-coordinate velocity
and a part that is in general non-linear in the past reaction coordinates but does not depend on velocities.
Our hybrid scheme thus combines all desirable properties of the Zwanzig and Mori projection schemes. 
%One memory kernel is a function of time only while the other also accounts for spatial dependencies. 
% The spatially dependent memory term represents a correction to the widely used ad-hoc GLE which consists of a PMF and a memory function that explicitly depends on time only. 
The non-linear memory friction contribution is shown to be related to correlations between the reaction-coordinate velocity and the random force.
We present a numerical method to compute all parameters of our GLE,
in particular the non-linear memory  friction function and the random force distribution, from a trajectory in reaction coordinate space.
%In this way, it is possible to test whether or not it is justified to neglect spatial memory effects in the presence of a non-linear potential. 
%Further, the introduced algorithm enables a sampling of the random force term in the GLE, 
%allowing the study of its statistical properties from a single trajectory, even in the presence of a PMF. 
We apply our method on the dihedral-angle dynamics of a butane molecule in water obtained from atomistic molecular dynamics simulations. For this example, we  demonstrate that  non-linear memory friction is present and that the random force exhibits significant non-Gaussian corrections.
%The algorithm is tested on two exactly solvable simple models. One where local effects can be neglected, and one where they have to be included.
We also present the derivation of the GLE for multidimensional reaction coordinates that are general functions of all positions in the phase space of the underlying many-body system; this  corresponds to a systematic coarse-graining procedure that preserves not only the correct equilibrium behavior but also the correct dynamics of the coarse-grained system.
\end{abstract}

% insert suggested keywords - APS authors don't need toergebnisse erhalten do this
%\keywords{Statistical Physics, Non-Markovian Effects, Spatially Dependent Memory}

\maketitle

%Popular Summary: Physical Review X requires authors to submit a nontechnical summary that conveys the context, the essential message(s), and the significance of the work to all readers. The summary should be concise (approximately 250 words), readable, objective, and have broad appeal. Please avoid including mathematical expressions.

\section{Introduction}
Most interesting physical systems are interacting many-body systems. 
When dealing with the kinetics of such systems, one is typically interested in the dynamics of a low-dimensional reaction coordinate, which is, however, generally influenced by the entire system \cite{van_kampen_remarks_1998}. 
Examples include the motion of a particle in a liquid \cite{espanol_force_1993,bocquet_brownian_1994,zwanzig_hydrodynamic_1970,franosch_resonances_2011,lesnicki_molecular_2016, daldrop_external_2017}, vibrational modes of a molecule in the gas phase or in a liquid \cite{straub_calculation_1987,berne_dynamic_1990,tuckerman_vibrational_1993,gottwald_applicability_2015},
chemical or associative reactions between molecules \cite{adelman_generalized_1980,ciccotti_derivation_1981,guardia_generalized_1985,bocquet_friction_1997,canales_generalized_1998} 
and protein folding \cite{plotkin_non-markovian_1998,medina_transition_2018,satija_generalized_2019,ayaz_non-markovian_2021}.
To predict the dynamics of the reaction coordinate, one in principle 
has to solve the equation of motion of  the underlying many-body system,
which is typically analytically impossible and is only numerically possible for small systems
and over short times. 
The  very attractive idea of coarse-grained modeling is to replace the description of the full many-body system 
by a description in terms of the reaction coordinates only. The challenge is to 
derive the appropriate equation of motion that describes the dynamics of the reaction coordinates accurately while maintaining numerical efficiency.
%would accurately and numerically efficiently describe the dynamics of  the reaction coordinates.
For some biologically relevant scenarios, such as the folding of a protein, 
sufficiently long simulations of the full system dynamics can be performed \cite{lindorff_fast_2011,best_native_2013,chung_structural_2015}, 
but even for these cases, the interpretation of the results typically  requires mapping onto a low-dimensional reaction coordinate. 

Rigorous coarse-graining methods based on projection operator techniques 
were introduced by Zwanzig and Mori, 
which are directly applied to the Liouville equation that describes the  dynamics of a classical many-body system  governed  by a  time-independent Hamiltonian~\cite{zwanzig_memory_1961,mori_transport_1965}
(in fact,  a similar approach applicable to quantum systems
was developed by Nakajima even earlier \cite{nakajima_quantum_1958}).
The result of the projection is a coarse-grained equation of motion for the chosen set of reaction coordinates, the so-called 
generalized Langevin equation (GLE). 
It contains three distinct terms: a force term  due to a
potential that depends on the reaction coordinates,  a  memory friction contribution
that involves the past time dynamics of the reaction coordinates,
and a  force that explicitly  depends on the initial state of the entire many-body system 
and  which is typically interpreted as a random or stochastic force. 
The GLE is therefore an integro-differential stochastic equation.
% that for most systems of interest is non-linear. 
It should be noted that  the Zwanzig and Mori  projection schemes give rise to fundamentally 
different GLEs for non-linear systems, which are both rigorous  and reproduce the system dynamics described by the reaction coordinates
exactly~\cite{zwanzig_memory_1961,mori_transport_1965}. 
However,  except a few notable exceptions \cite{grabert_microdynamics_1980,chorin_optimal_2000,kinjo_equation_2007, hijon_morizwanzig_2010, meyer_non-stationary_2017,herrera_tractable_2020},
the exact Zwanzig or Mori equation have rarely  been used in practice for non-trivial, i.e. non-linear, systems,
for different reasons: In the Mori framework,  the force from the potential as well as the memory friction 
are linear  in the reaction coordinate and their velocities, respectively,  and therefore 
all non-linearities are accounted for by the random force, which thus becomes non-Gaussian and is difficult 
to parameterize;
 in the Zwanzig framework, 
 the potential term in the GLE  is in general non-linear and corresponds to the potential of mean force (PMF), which
 ensures the correct equilibrium distribution of the reaction coordinates \cite{chorin_optimal_2000}, 
 which is a desired property. On the other hand, the memory friction is a general function  of both the reaction coordinates  and their velocities, 
 which poses severe problems when
estimating  such a function from simulation or experimental data. 

As a consequence,  many previous works considered 
a simplified form of the GLE, which  in this paper we refer to  as the approximate GLE. It   contains the non-linear PMF
and a  memory friction   that is linear in the
velocity of the reaction coordinate  \cite{darve_numerical_2006,lange_collective_2006,carof_two_2014, lesnicki_molecular_2016, jung_iterative_2017, daldrop_butane_2018, lee_multi_2019, klippenstein_introducing_2021,vroylandt_likelihood_2021}.
 In principle, this approximate GLE follows from the Zwanzig GLE, assuming that the friction memory  depends only linearly  on the past
reaction coordinates and is independent of   the  reaction-coordinate velocities. 
The validity of this approximation can typically not be checked in a systematic manner.
The applications of the approximate GLE range from non-Markovian rate theory \cite{grote_stable_1980, hanggi_thermally_1982, pollak_theory_1989}, over protein folding dynamics \cite{plotkin_non-markovian_1998, medina_transition_2018, satija_generalized_2019, ayaz_non-markovian_2021} to molecular diffusion
and conformational dynamics \cite{lesnicki_molecular_2016,daldrop_external_2017, daldrop_butane_2018,kappler_cyclization_2019}. 
Methods to derive memory functions from trajectory data for non-linear systems  within the framework of the approximate
 GLE have been introduced 
and it was demonstrated that the resulting GLE  correctly describes the multi-scale fractal dynamics of protein folding 
~\cite{ayaz_non-markovian_2021} and the vibrational spectra of molecules in non-linear bond-length and
bond-angle potentials \cite{brunig_proton_2021}.
%The ad-hoc GLE is a practical tool to incorporate non-Markovianity in theoretical models and was utilized so to develop influential theories such as the Grote-Hynes theory of reaction rates. 
Although widely used, the validity of the approximate
GLE in the presence of a non-linear potential is subject to ongoing discussions \cite{klippenstein_cross-correlation_2021,glatzel_interplay_2021}.
%Deriving such an equation for a low dimensional reaction coordinate of choice where the equation has no dependency on the non-relevant part of the system is often referred to as the closure problem. 
%Neglecting spatial dependencies in the memory term relies on certain assumptions about the stochastic term of the GLE. These assumptions are made because the exact forms obtained otherwise are difficult to deal with, and a coarse-grained model, which is both exact and computationally manageable, is lacking for most relevant applications. 

In this paper, we introduce a projection method that  is a hybrid of the Zwanzig and Mori projection schemes.
As an advantage over the Mori projection scheme, the resulting GLE contains the force stemming from the  generally  non-linear PMF, 
which by itself guides the system into the correct equilibrium distribution in the long-time limit.
As an advantage over the Zwanzig projection scheme, the generally non-linear  memory friction  does not depend on the velocity of the reaction coordinate but only on the
reaction coordinate itself, which significantly simplifies the numerical estimation of the memory function from trajectory data.
We develop the necessary framework to compute all parameters of  the resulting GLE from  trajectories of a reaction coordinate.
Thus, we present  data-based methods
i) to derive  the non-linear   memory friction from simulation or experimental  trajectories,
ii) to thereby examine  the validity of the approximate GLE,  and
iii) to  study the distribution and correlation of the random force from  trajectories.
We also derive a multidimensional GLE in terms of a general set of reaction coordinates that are arbitrary functions of the positions
of the underlying many-body system; this  constitutes a rigorous derivation of the equations of motion that accurately describe the 
equilibrium and dynamic behavior of coarse-grained systems. 
For the explicit example of the 
dihedral-angle dynamics of a butane molecule in water, obtained from atomistic molecular dynamics simulations,
we  demonstrate that  non-linear memory  friction  is present and that the random force exhibits significant   non-Gaussian corrections.
Therefore, we find that even for this simple molecular system, the approximate
 GLE, which neglects non-linear memory  friction and assumes Gaussian random forces, 
does not correctly describe the dynamics.

% formulated by Nakajima \cite{nakajima_quantum_1958}, Mori \cite{mori_transport_1965}, Zwanzig \cite{zwanzig_memory_1961} and Grabert \cite{grabert_microdynamics_1980} in the context of nonequilibrium statistical mechanics. The projection formalism results in an exact decomposition of the equations of motion for some set of relevant observables into three parts; two parts depend on the relevant observables alone, the third term depends on the initial state of the full system. The latter is modeled as a stochastic process to make the decomposition a practical computational tool \cite{hijon_morizwanzig_2010}. For large systems, this is reasonable since it is a function of an immense number of variables and thus, represents a fluctuating force. The form of the first two terms in the decomposition depends on the choice of projection.\\

The paper is organized as follows: First, we introduce the Hamiltonian of the many-body system, as well as our notation, and we present important expressions for correlation functions
and conditional averages.
We then review  the Mori and Zwanzig projection schemes  and highlight practical problems of  the resulting GLEs. 
After this, we introduce our hybrid projection scheme and derive the  GLE that features a non-linear PMF 
and non-linear memory friction.  
In the subsequent section, we  introduce an  algorithm to extract all parameters of our GLE from trajectories. 
In the final section, we apply our formalism  on  two exactly solvable  model systems  and on MD trajectories for  the dihedral angle dynamics of a butane molecule in water.

\section{Hamiltonian Model, Notation and Useful Properties\label{sec_definitions}}
We denote the phase space of a system of $N$ interacting particles in three-dimensional space  by $\Omega$. 
%$\Omega$ consists of all microstates that the system can be in. 
One specific microstate, i.e., a point in $\Omega$, is denoted by $\omega = (\ve R, \ve P) = (\ve r_1, \ve r_2,\dots,\ve r_N, \ve p_1, \ve p_2, \dots, \ve p_N)$ which is a $6N$ vector of the Cartesian positions $\ve r_i=(r^x_{i}\; r^y_{i}\; r^z_{i})$, and the  conjugate momenta $\ve p_i = (p^x_{i}\; p^y_{i}\; p^z_{i})$ of all $i=1,2,\dots,N$ particles in the system. 
%For the sake of clarity, we omit the time dependency in the initial phase space coordinates, e.g., we use $\ve r_n(0)\equiv\ve r_n$. 
The Hamiltonian of the system is an invariant of motion and splits into a kinetic and a potential part
\begin{align}
\label{eq_Hamiltonian}
H(\omega) &= \sum_{i=1}^N \frac{\ve{p}_i^2}{2 m_i} + V(\ve R).
\end{align}
%As mentioned in the introduction, the form of the Hamiltonian in equation~\eqref{eq_Hamiltonian} is motivated by force-fields used in MD simulations.
The potential $V(\ve R)$ contains all interactions between the particles and possible external potentials. The only assumption on $V$ is that it is a function of the positions $\ve R$ only. The time evolution of a point $\omega$ in phase space is determined by Hamilton's equation of motion,
 which can be written in the form
\begin{align}
\label{eq_time_evolution}
\dot{\omega}_t &= L \omega_t,
\end{align}
where $\omega_t$ is the location of the system in phase space at time $t$ 
and $\dot{\omega}_t$  denotes the corresponding velocity, 
given the system was initially at $\omega_0$. For the sake of compact notation, 
we denote time dependencies of phase space coordinates by a subscript.
% and from now on, $\omega_0$ denotes the initial state of the complete system.
In eq.~\eqref{eq_time_evolution}, $L$ is the Liouville operator given by
\begin{align}
\label{eq_Liouville}
L &= \sum_{n=1}^N\left( \frac{\ve p_n}{m_n}\cdot \nabla_{\ve r_n} - \left(\nabla_{\ve r_n} V(\ve R) \right)\cdot \nabla_{\ve p_n} \right).
\end{align}
All of the operators that we consider in this work, including the Liouville operator $L$, act on the initial phase space position $\omega_0$. 
From eq.~\eqref{eq_time_evolution}, it follows that the system is propagated in time by the operator $e^{tL}$, i.e., $e^{tL}\omega_0 = \omega_t$.
%\begin{subequations}
%\begin{align}
%\Phi(t): \Omega &\to \Omega\\
%\omega_0 &\mapsto \omega_t\\
%\Phi(t) &= e^{tL}.
%\end{align}
%\end{subequations}
We consider observables that are real-valued functions of phase-space coordinates only and that depend on time implicitly via the 
time dependence of a trajectory moving  in phase space. For the sake of notational brevity, we also denote the time dependency of  observables by a subscript too, i.e., $A_t \equiv A(\omega_t)=A(\omega_0, t)  $.
%\begin{subequations}
%\begin{align}
%A_i: \Omega &\to \mathbb{R},\\
%\omega_t &\mapsto A_i(\omega_t)=A_i(\omega_0,t).
%\end{align}
%\end{subequations}
Using the chain rule for differentiation, it follows that the time evolution of an observable $A_t$ is also governed by the Liouville equation \cite{zwanzig_nonequilibrium_2001}
\begin{align}
\label{eq_Liouville_observable}
\dot{A}_t &= L A_t,
\end{align}
where $\dot{A}_t$ denotes the time derivative of $A_t$.
Thus, the time propagation operator of an observable in the initial state $A(\omega_0)\equiv A_0$, is also given by $e^{tL}$. From this, it follows that
\begin{align}
\label{eq_propagator_prob}
A(\omega_{t+t^\prime})=e^{(t+t^\prime)L}A(\omega_0) = e^{tL}A(\omega_{t^\prime}) = A(\omega_{t^\prime},t).
\end{align}
Eq.~\eqref{eq_propagator_prob} describes how observables are propagated in time by $e^{tL}$ and will be used throughout our derivations.
All observables  are elements of a Hilbert space, i.e., a vector space equipped with an inner product. Let $A$ and $B$ denote two system observables. 
For the inner product, we choose
\begin{align}
\label{eq_inner_product}
\langle A_t , B_{t^\prime} \rangle  \equiv  \int_\Omega \mathrm{d}\omega_0\, \rho_\mathrm{eq}(\omega_0) A(\omega_0,t) B(\omega_0,t^\prime),
\end{align}
where $\rho_\mathrm{eq}(\omega_0) = e^{-\beta H(\omega_0)}/Z$ is the canonical Boltzmann distribution with the inverse thermal energy $\beta = 1/k_BT$ and the partition function $Z = \int_\Omega\mathrm{d}\omega_0\, e^{-\beta H(\omega_0)}$. The inner product in eq.~\eqref{eq_inner_product} thus corresponds to an equilibrium time correlation function which establishes the link to statistical mechanics. 
The  average of a single  observable $B_t$ is  given by $\langle B_t\rangle  \equiv  \langle B_t,1\rangle  $ and does not depend on time.
Because of the form of the Hamiltonian in eq.~\eqref{eq_Hamiltonian}, the Boltzmann distribution factorizes into a position and a momentum-dependent part
\begin{align}
\rho_\mathrm{eq}(\omega_0) &= \frac{1}{Z}e^{-\beta H(\omega_0)} = \rho_\mathrm{kin}(\mathbf{P}_0)\, \rho_\mathrm{pot}(\mathbf{R}_0),
\end{align}
where $\rho_\mathrm{kin}(\mathbf{P}_0) \propto \prod_{i=1}^N \exp\left(-\beta\,\ve p_{i,0}^2/2 m_i\right)$ is a Gaussian with zero mean. 
%At this point, we would like to emphasize the importance of this property. It will be used again and again in the course of this work. 
With respect to the inner product in eq.~\eqref{eq_inner_product}, the Liouville operator, as defined in eq.~\eqref{eq_Liouville}, is anti-self-adjoint \cite{zwanzig_nonequilibrium_2001}
\begin{align}
\label{eq_Liouville_anti_sa}
\langle L A_t, B_{t^\prime} \rangle &= - \langle A_t, L B_{t^\prime} \rangle.
\end{align}

\subsection{Conditional Averages}
In addition to time-correlation functions calculated over the entire phase space $\Omega$, as in eq.~\eqref{eq_inner_product}, 
we will also use conditional time-correlation functions that result from averages over a hyper surface in phase space on which an observable of choice at the initial time $t=0$, $A_0=A(\hat{\omega}_0)$, takes a constant value $A(\omega_s)$. A conditional correlation of two observables 
$B_t=B(\omega_t)=B(\omega_0,t)$ and $C_{t^\prime}=C(\omega_{t^\prime})=C(\omega_0,t^\prime)$ is defined by \cite{grabert_microdynamics_1980, chorin_optimal_2000}
\begin{equation}
\label{eq_conditional_avrg}
\langle B_t, C_{t^\prime}\rangle_{A_s} = 
\frac{\langle\delta\left[A(\widehat{\omega}_0)-A(\omega_s)\right], B(\widehat\omega_0,t) C(\widehat\omega_0,t^\prime)\rangle}{\langle\delta[A(\widehat{\omega}_0)-A(\omega_s)]\rangle}.
\end{equation}
In eq.~\eqref{eq_conditional_avrg}, the phase space variable with a hat,  $\widehat{\omega}_0$,  is integrated over. The phase space variable $\omega_s$ is not.
 %\textcolor{red}{, i.e., $A(\omega_s)$ is the value that determines the hyper-surface of the} conditional correlation. 
 Therefore, $\langle B_t, C_{t^\prime}\rangle_{A_s}$ is a function of $\omega_s$, and the times $t$ and $t^\prime$.
The conditional average of a single  observable $B_t$ is  given by $\langle B_t\rangle_{A_s} \equiv  \langle B_t,1\rangle_{A_s}  $.

Finally, we give a few relations which will be frequently used later on. 
We repeat that a conditional average is a function of phase space via the conditional function $A_s =
A(\omega_s)$ in eq.~\eqref{eq_conditional_avrg}. The time propagation of a conditional average is thus given by
\begin{align}
\label{eq_propagation_cond_avrg}
e^{tL}\langle B_{t^\prime}\rangle_{A_0}=\langle B_{t^\prime}\rangle_{A_t}.
\end{align}
The  normalized probability that an observable $A_t$ has the value $a$ is given by  $\mathbb{P}(a) \equiv \langle \delta(A_t-a)\rangle$,
from which the potential of mean force (PMF) for an observable follows as  \cite{darve_numerical_2006}
\begin{align}
\label{eq_def_PMF}
U_\mathrm{PMF}(a) &\equiv  -k_BT \ln\mathbb{P}(a).
\end{align}
Acting with the Liouville operator on a delta function gives \cite{hijon_morizwanzig_2010}
\begin{align}
\label{eq_Liouville_on_delta}
L\delta(A_t-a) = -\dot{A}_t \frac{\mathrm{d}}{\mathrm{d}a}\delta(A_t-a).
\end{align}
Using the definition in eq.~\eqref{eq_conditional_avrg} together with the relations in eq.~\eqref{eq_Liouville_anti_sa}, eq.~\eqref{eq_Liouville_on_delta} and the PMF defined  in eq.~\eqref{eq_def_PMF}, we derive in appendix~\ref{sec_App_eq13}  the important relation \cite{hijon_morizwanzig_2010}
\begin{align}
\label{eq_cond_avrg_rel}
\langle LB_{t^\prime}\rangle_{A_t} = \frac{\mathrm{d}}{\mathrm{d}A_t}\langle \dot{A}_0,B_{t^\prime}\rangle_{A_t}-\beta\langle \dot{A}_0,B_{t^\prime}\rangle_{A_t}\frac{\mathrm{d}}{\mathrm{d}A_t}U_\mathrm{PMF}(A_t).
\end{align}

\section{\label{sec_POM} Projection Operator Method}
We  now derive the equation of motion for an arbitrary scalar observable $A_t$, which can of course also be the position of a single particle \cite{zwanzig_nonequilibrium_2001}. 
The derivation for a general multi-dimensional observable is given in appendix~\ref{sec_App_multidim}.
A projection $P$ is a linear, idempotent operator, i.e., for arbitrary scalars $c_1, c_2$, it fulfills the properties $P(c_1 A_t + c_2 B_t)=c_1 PA_t+ c_2 PB_t$ and $P^2 = P$. The operator $Q=1-P$ projects onto the \textit{complementary} subspace with $1$ being the identity operator. For briefness, we will refer to the subspace onto which $P$ projects as the \textit{relevant} subspace. The operators $P$ and $Q$ can be used to decompose the Liouville equation $\ddot{A}_t=L\dot{A}_t$ for the observable $\dot{A}_t$ as
\begin{align}
\label{eq_Liouville_decompose}
\ddot{A}_t &= e^{tL} (P + Q) L \dot{A}_0 = e^{tL} P L \dot{A}_0 + e^{tL} Q L \dot A_0.
\end{align}
To obtain an equation of motion for $A_t$ from eq.~\eqref{eq_Liouville_decompose}, we introduce the operator 
\begin{align}
\label{eq_phi}
\Phi(t) &= e^{tL}Q.
\end{align}
$\Phi(t)$ propagates  the part of an observable that lies in the complementary subspace in time. For $\Phi(t)$ we find
\begin{subequations}
\begin{align}
\frac{\mathrm{d}}{\mathrm{d}t}e^{tL}Q &= e^{tL}LQ = e^{tL}QLQ + e^{tL}PLQ,\\
\dot{\Phi}(t) &= \Phi(t)LQ + e^{tL}PLQ.\label{eq_deq_propagator}
\end{align}
\end{subequations}
Eq.~\eqref{eq_deq_propagator} is an inhomogenous differential equation  of first order. Using $\Phi(0) = Q$, as follows from eq.~\eqref{eq_phi}, the solution reads
\begin{align}
\Phi(t) &= Q e^{tLQ} + \int_0^t\mathrm{d}u\,e^{uL}PLQe^{(t-u)LQ}.\label{eq_phi_propagator}
\end{align}
By using $Q e^{tLQ} = e^{tQL}Q$ and the substitution $s=t-u$ in eq.~\eqref{eq_phi_propagator}, we find
\begin{align}
\Phi(t) &= e^{tL}Q = e^{tQL}Q + \int_0^t\mathrm{d}s\,e^{(t-s)L}PL e^{sQL}Q.\label{eq_sol_propagator}
\end{align}
Since the operator $e^{tQL}Q$ exhibits a $Q$ operator on the left side when the exponential is expanded, the first term on the r.h.s. of eq.~\eqref{eq_sol_propagator} stays in the complementary subspace for all times. The second term describes the effect 
of the complementary subspace on  the relevant subspace. By factoring out the operator $Q$ on the r.h.s. of eq.~\eqref{eq_sol_propagator}, one obtains the \textit{Dyson decomposition} \cite{dyson_radiation_1949, feynman_operator_1951, evans_statistical_2008} of the propagator $e^{tL}$
\begin{align}
e^{tL} = e^{tQL} + \int_0^t\mathrm{d}s\,e^{(t-s)L}PL e^{sQL}.\label{eq_dyson_decomposition}
\end{align}
Replacing $e^{tL}Q$ in eq.~\eqref{eq_Liouville_decompose} by eq.~\eqref{eq_sol_propagator} leads to the GLE  for $A_t$ in terms of a general projection $P$ \cite{zwanzig_memory_1961, mori_transport_1965, zwanzig_nonequilibrium_2001}
\begin{subequations}
\begin{align}
\label{eq_GLE_1}
&\ddot{A}_t = e^{tL}PL \dot A_0 + \int_0^t\mathrm{d}s\,e^{(t-s)L}PL\,F^R(s) + F^R(t),\\
\label{eq_F_operator}
&F^R(t) \equiv  e^{tQL}QL \dot A_0 = Q e^{t LQ} L \dot A_0.
\end{align}
\end{subequations}
The function $F^R(t)$ stays in the complementary subspace for all times and
 is an explicit  function of the initial state of the entire system, i.e., $F^R(t)=F^R(\omega_0,t)$. 
 Hence,  for large systems, it can be interpreted as a random or stochastic function. For the sake of brevity, we will write out  the $\omega_0$ dependence of $F^R(t)$ only when it improves clarity. The first term on the r.h.s. of eq.~\eqref{eq_GLE_1} represents the time evolution of the part of 
 $\ddot{A}_0=L \dot{A}_0$ which  lies in the relevant subspace and reflects a deterministic  force. 
 The second term on the r.h.s. of eq.~\eqref{eq_GLE_1} 
 is due to the relevant part of $L F^R(\omega_0,t)$ and describes dissipative effects.
Clearly, the explicit form of eq.~\eqref{eq_GLE_1} depends on the specific form of the projection operator $P$. Before we introduce our hybrid
projection scheme, 
we will  present the GLE's generated by the Mori projection $P_M$ and by the Zwanzig projection $P_Z$.

\subsection{ Mori Projection}
The Mori projection applied on an observable $A_t$ is given by \cite{mori_transport_1965}
\begin{align}
\label{eq_mori_projection}
P_M A_t = \frac{\langle A_t, B_{0}\rangle}{\langle B_{0}^2 \rangle}B_{0} + \frac{\langle A_t, \dot{B}_{0}\rangle}{\langle \dot{B}_{0}^2 \rangle}\dot{B}_{0},
\end{align}
and uses the inner product defined  in eq.~\eqref{eq_inner_product}. The observables one projects onto, 
i.e., $B_0$ and $\dot{B}_0$, are referred to as the \textit{projection functions}. 
The projection in eq.~\eqref{eq_mori_projection} maps any observable $A_t$ onto the subspace of all functions linear in the observables $B_0$ and $\dot{B}_0$. In addition to being linear and idempotent, $P_M$ is self-adjoint w.r.t. to the inner product in eq.~\eqref{eq_inner_product}, i.e., for two arbitrary observables $A_t, C_{t^\prime}$, the relation
\begin{align}
\langle P_M A_t, C_{t^\prime}\rangle = \langle A_t, P_M C_{t^\prime}\rangle\label{eq_projection_orthogonal}
\end{align}
holds. Thus, it is an orthogonal projection, since all functions $P_M A_t$ and $Q_M C_{t^\prime}$ are orthogonal, i.e.,
\begin{align}
\langle P_M A_t, Q_M C_{t^\prime}\rangle = 0,
\end{align}
as follows directly from eq.~\eqref{eq_projection_orthogonal} and from the idempotence of $P$. For $P = P_M$ and choosing the projection functions to be $B_t = A_t$ and $\dot{B}_t=\dot{A}_t$, i.e., projecting onto the observable of interest itself, eq.~\eqref{eq_GLE_1} takes the form \cite{mori_transport_1965, zwanzig_nonequilibrium_2001}
\begin{subequations}
\label{eq_mori_GLE}
\begin{align}
\ddot{A}_t &= -\mathrm{K}\, A_t - \int_0^t\mathrm{d}s\,\Gamma^M(s) \dot A_{t-s} + F^R(\omega_0,t),\label{eq_mori_p}\\
\mathrm K &= \frac{\langle \dot{A}_{0}^2\rangle}{\langle A_{0}^2\rangle},
 \quad \Gamma^M(t) = \frac{\langle F^R(\omega_0,t), F^R(\omega_0,0) \rangle}{\langle \dot A_0^2\rangle},
 \label{eq_mori_memory}
\end{align}
\end{subequations}
where $\Gamma^M(s)$ is the memory friction kernel  obtained from the Mori projection. Eq.~\eqref{eq_mori_GLE} is an exact decomposition of the Liouville equation into three terms: the first term is a generalized  force due to a potential of quadratic form; the second term accounts for linear friction and includes the memory kernel $\Gamma^M(s)$,
  which is related via eq.~\eqref{eq_mori_memory} to the second moment of the random force $F^R(\omega_0,t)$, 
defined in eq.~\eqref{eq_F_operator}. The exact form of the memory function can only be computed for very simple models,
%, as will be demonstrated later. 
for realistic systems and practical applications it is infeasible to compute  since the fluctuating term $F^R(\omega_0,t)$ is an explicit 
 function of the initial state of the entire system. Instead, one typically models the function $F^R(t)$ as a stochastic process with zero mean and a second moment given in eq.~\eqref{eq_mori_memory}. Although information on higher-order moments of $F^R(t)$ can be obtained from the 
 Mori formalism,  $F^R(t)$  is typically  assumed to be Gaussian. In general, however, this assumption can not hold, 
 since $F^R(t)$ contains all non-linearities that $A_t$ may exhibit. 
 Thus, imposing $F^R(t)$ to be a Gaussian variable  becomes a bad approximation for non-linear systems,
 %In fact, from the formalism, it does not even follow that the equilibrium average of $F^R(t)$ vanishes. 
 which reflects a fundamental short-coming of the Mori projection scheme for practical applications.
 %\textcolor{red}{and is the reason why the Mori GLE is not suitable for the stochastic modeling of dynamics in the presence of non-quadratic potentials.}
%For this reason, $F^R(t)$ should contain as little information as possible so that its impact on the statistics is as little as possible. This is the primary motivation to introduce a non-linear potential term into the GLE that accounts for non-linearities in $A_t$ and, further, ensures that the mean behavior of the reduced dynamics, as described by the GLE, follows the mean dynamics of the complete system \cite{chorin_optimal_2000}. 
%Before we continue with the Zwanzig projection, it is worth mentioning that the Mori formalism can be extended to systems far from equilibrium. By using a time-dependent distribution instead of the Boltzmann distribution in eq.~\eqref{eq_inner_product}, one can obtain a non-equilibrium GLE with a structure similar to equation~\eqref{eq_mori_GLE} \cite{meyer_non-stationary_2017, meyer_non-markovian_2020}.

\subsection{Zwanzig Projection}
Contrary to the Mori projection, the Zwanzig projection $P_Z$ of an observable $A_t$ is non-linear in the projection functions 
$B_0$ and $\dot{B}_0$ \cite{zwanzig_memory_1961}
\begin{align}
\label{eq_projection_zwanzig}
P_Z A_t &= \frac{\langle\delta\left(B(\widehat{\omega}_0)-B(\omega_0)\right)\delta(\dot{B}(\widehat{\omega}_0)-\dot{B}(\omega_0)), 
A(\widehat{\omega}_0,t) \rangle}{\langle\delta\left(B(\widehat{\omega}_0)-B(\omega_0)\right)\rangle\langle \delta(\dot B(\widehat{\omega}_0)-\dot B(\omega_0)) \rangle}\nonumber\\
 &= \langle A_t\rangle_{B_0, \dot B_0},
\end{align}
where we repeat  that phase-space variables with a hat inside inner products, i.e., $\widehat{\omega}_0$, are integrated over. 
The Zwanzig projection thus is a conditional average as defined in eq.~\eqref{eq_conditional_avrg} and is 
 linear, idempotent and self-adjoint, similar to the Mori projection.
The resulting GLE from the Zwanzig projection is best illustrated
 by choosing the observable of interest  to be the momentum of a single particle,
  $\dot{A}_0=\mathbf{p}_0$,
  and  the projection functions as the position and the linear momentum of the same  particle, i.e., 
 $B_0\to \ve r_0$, $\dot{B}_0\to\ve p_0$. 
With this, eq.~\eqref{eq_GLE_1} becomes \cite{darve_computing_2009}
\begin{subequations}
\begin{align}
\label{eq_GLE_zwanzig}
\dot{\ve p}_t &= -\nabla_{\ve r_t}  U_\mathrm{PMF}(\ve r_t) + \ve F^R(\omega_0,t) \nonumber \\
& \quad + \int_0^t\mathrm{d}s\left[\left( \frac{\nabla_{p_{s}}}{ \beta}- \frac{  \ve p_{s}}{m} \right)\right]^\mathrm{T}\cdot \Gamma^Z(t-s, \ve r_{s}, \ve p_{s})
\end{align}
\end{subequations}
with a memory friction kernel defined by
\begin{equation}
\beta \Gamma^Z_{ij}(t-s,\ve r_s, \ve p_s) =  \langle F_i^R(0), F_j^R(t-s)\rangle_{\ve r_s, \ve p_s}.
\end{equation}
Here, $U_\mathrm{PMF}(\ve r) = -k_BT\ln\langle \delta(\ve r_0-\ve r)\rangle$ denotes the potential of mean force (PMF) defined in eq.~\eqref{eq_def_PMF},
which creates  in the GLE a force  on the  particle that tends to establish the equilibrium positional distribution. 
This is the main advantage over the  Mori projection, since 
this ensures the correct equilibrium   behavior once we switch to a stochastic description and replace the fluctuating force $F^R(t)$ by a Gaussian stochastic variable with zero mean \cite{chorin_optimal_2000}. 
The memory friction kernel $\Gamma^Z$  is a $3\times 3$  matrix  that,  as a result of the conditional average,  is a function of particle position  $\ve r_s$
and particle momentum $ \ve p_s$.  This is the main
drawback of the GLE  in eq.~\eqref{eq_GLE_zwanzig}, since  the position and momentum dependence is difficult to deal with in applications. 
As a way out, one typically invokes  the ad-hoc assumption that the memory function  is   
independent of  position and momentum, i.e., $\Gamma^Z(t-s,\mathbf{r}_s, \mathbf{p}_s)\approx \Gamma^\mathrm{app}(t-s)$. 
This assumption leads to an approximate GLE  that  is amply used in  literature \cite{grote_stable_1980,hanggi_thermally_1982,pollak_theory_1989,plotkin_non-markovian_1998,darve_numerical_2006,lange_collective_2006,jung_iterative_2017,daldrop_external_2017,daldrop_butane_2018,medina_transition_2018,satija_generalized_2019, lee_multi_2019,klippenstein_introducing_2021,vroylandt_likelihood_2021,ayaz_non-markovian_2021,brunig_proton_2021} 
and reads
\begin{align}
\label{eq_GLE_adhoc}
\dot{\ve p}_t &= -\nabla_{\ve r_t}   U_\mathrm{PMF}(\ve r_t)  - \int_0^t\mathrm{d}s\,\Gamma^\mathrm{app}(t-s)\cdot \frac{\ve p_{s}}{m} + \ve F^R(t).
\end{align}
While for  various applications the approximate  GLE has been demonstrated to reproduce the full system dynamics very accurately 
\cite{ayaz_non-markovian_2021, brunig_proton_2021},
it is difficult to check  for realistic systems whether  the ad-hoc assumption 
$\Gamma^Z(t-s,\mathbf{r}_s, \mathbf{p}_s)\approx \Gamma^\mathrm{app}(t-s)$ is in fact valid.  
 This is one motivation for our hybrid projection scheme, since it allows to derive all parameters of 
the exact GLE from trajectory data and thereby to access the validity of the approximate  GLE explicitly. 

\section{Hybrid GLE \label{sec_derivation}}
Our  projection operator  $P_H$ is a hybrid of the Mori and Zwanzig projection operators
%After we have introduced the projection operator, we will show that it is an orthogonal projection, i.e., it fulfills $\langle P A_i(t), A_j(t^\prime)\rangle = \langle A_i(t), P A_j(t^\prime)\rangle$. Using this, we will  derive a GLE where the memory matrix only depends on time, but the PMF is still present.
and  is written in the form $P_H = P_x + P_p$. Here, we derive the GLE for a scalar observable $A_t=A(\omega_t)$,
the derivation for a general multi-dimensional observable is given in appendix~\ref{sec_App_multidim}.
Using general projection functions $B_0 = B(\ve R_0)$, which is  a function of  positions only,
and $\dot B_0= \dot B(\ve R_0, \ve P_0)$, which in general is  a function of  positions and momenta,
 the hybrid projection operator  is given by
\begin{subequations}
\label{eq_projection}
\begin{align}
P_H A_t &= P_x A_t + P_p A_t,\\
P_x A_t&= \frac{\langle \delta(B(\widehat{\mathbf{R}}_0)-B(\mathbf{R}_0)), A(\widehat{\omega}_0,t) \rangle}{\langle \delta(B(\widehat{\mathbf{R}}_0)-B(\mathbf{R}_0)) \rangle} = \langle A_t\rangle_{B_0},\label{eq_projection_x}\\
P_p A_t &= \frac{\langle \dot B_0, A_t\rangle}{\langle \dot B_{0}^2\rangle}\,\dot B_{0}.\label{eq_projection_p}
\end{align}
\end{subequations}
The projection $P_x$ is a conditional average, defined  in eq.~\eqref{eq_conditional_avrg}, onto the  observable $B_0 = B(\ve R_0)$, which
 is a function of positions $\ve R$ only. As a result, the conditional average is independent of momenta.   
 In appendix~\ref{sec_App_idem} we show that $P_xP_p = P_pP_x = 0$,
from which follows  that $P_H^2 = P_H$, so that  $P_H$ is idempotent in addition to being linear  and hence is a projection. 
In appendix~\ref{sec_App_ortho}, we show that $P_H$ is self-adjoint w.r.t the inner product defined in eq.~\eqref{eq_inner_product}, i.e., it fulfills the 
property in eq.~\eqref{eq_projection_orthogonal}.
Therefore, $P_H$ is an orthogonal projection. Again, we denote the projection onto the complementary subspace of $P_H$ by $Q_H = 1 - P_H$, where $1$ is the identity operator.
In appendix~\ref{sec_App_equil_avrg}, we prove for the projections $P_H$, $Q_H$ of an arbitrary observable $A_t$ the  important property
\begin{align}
\label{eq_avereage_orthogonal_dyn}
\langle P_H A_t \rangle &= \langle A_t \rangle \Rightarrow \langle Q_H A_t \rangle = 0.
\end{align}
Hence, the equilibrium ensemble average of any observable that lies completely in the complementary subspace 
%\textcolor{red}{\sout{spanned by $Q_H$}} 
vanishes. As an important consequence, the random force
$F^R(t)$ defined  in eq.~\eqref{eq_GLE_1}  lies  completely in the complementary subspace for all times and, therefore, 
has  a vanishing equilibrium average. This property is also obtained for the Zwanzig projection, but not for the Mori projection.

%\subsection{The Potential of Mean Force}
In the remainder, we choose the observable of interest and  the projection function to coincide,  
$B(\mathbf{R}_t)=A(\mathbf{R}_t)$.
Therefore, the  GLE we derive from our hybrid scheme describes observables that are functions of positions only, 
such as the center of mass position, distances and angles. As an important property,
Our hybrid projection
 $P_H$ projects the observable $A_0$ and its velocity $\dot A_0$ onto themselves, meaning that
\begin{align}
P_H A_0 &= A_0, & P_H \dot A_0 &= \dot A_0.
\end{align}
With this choice for the projection function 
 and the specific form of the projection $P_H$ in eq.~\eqref{eq_projection}, we find for the first term on the r.h.s. of eq.~\eqref{eq_GLE_1},
\begin{subequations}
\label{eq_Upmf}
\begin{align}
&e^{tL} P_H L \dot A_0 = e^{tL}\left(P_x L \dot A_0 + P_p L \dot A_0\right),\label{eq_Upmf0}\\
&P_p L \dot A_0 \propto \langle \dot{A}_0, L\dot{A}_0 \rangle = -\langle L \dot{A}_0, \dot{A}_0 \rangle = 0,\label{eq_Upmf1}\\
&P_x L \dot A_0 = \langle L \dot A_0\rangle_{A_0}\nonumber\\
 & \quad \quad = \frac{\mathrm{d}}{\mathrm{d}A_0}\langle \dot{A}_0^2\rangle_{A_0}-\beta\langle \dot{A}_0^2\rangle_{A_0}\frac{\mathrm{d}}{\mathrm{d}A_0}U_\mathrm{PMF}(A_0),\label{eq_Upmf2}
\end{align}
\end{subequations}
where we used the relation in eq.~\eqref{eq_cond_avrg_rel} to obtain eq.~\eqref{eq_Upmf2}. 
Equation~\eqref{eq_Upmf2} describes the force due to a potential. To show this, we make use of the fact that the expectation value $\langle \dot A_0^2\rangle_{A_0}$ is strictly positive. Thus, we can  use it via
\begin{align}
\label{eq_gen_M}
\langle \dot A_0^2\rangle_{A_0} \equiv k_BT/M(A_0),
\end{align}
to define the generalized mass $M(A_0)$, which in general is a function of $A_0$. Using $M(A_0)$, eq.~\eqref{eq_Upmf2} can be simplified to
\begin{equation}
\label{eq_genU}
P_x L \dot A_0 =-\frac{1}{M(A_0)}\frac{\mathrm{d} U_\mathrm{eff}(A_0) }{\mathrm{d}A_0},
\end{equation}
where we defined the effective potential  as
\begin{equation}
\label{eq_gen_U1}
U_\mathrm{eff}(A_0) = U_\mathrm{PMF}(A_0) + k_BT\ln M(A_0).
\end{equation}
The effective potential  combines the effects of the PMF and the logarithmic effective mass.
%Since only the derivative of $\ln M(A)$ enters eq.~\eqref{eq_gen_U}, we rescale $M(A)$ inside the logarithm by an arbitrary constant to make it unitless so that the l.h.s. and r.h.s. of eq.~\eqref{eq_gen_U} have the same units. We choose this constant to be one (unit mass). 
%Applying the operator $e^{tL}$ onto eq.~\eqref{eq_gen_U1} propagates all observable $A_0$ to $A_t$, and we obtain the final form for the first term on the r.h.s. of eq.~\eqref{eq_GLE_1}. 

The second term on the r.h.s. of eq.~\eqref{eq_GLE_1} accounts for memory friction, the integrand 
for our hybrid projection reads  $e^{(t-s)L} (P_x + P_p) L F^R(s)$. 
The $P_p$ projection leads to a memory function of the same form as in the Mori projection
\begin{subequations}
\begin{align}
&P_p L F^R(s) = \frac{\langle \dot A_0, L F^R(s) \rangle}{\langle \dot{A}_0^2\rangle }\dot A_{0} = -\frac{\langle F^R(0), F^R(s) \rangle}{\langle \dot{A}_0^2\rangle}\dot A_{0},\\
\label{eq_Pp_memory}
&e^{(t-s)L} P_p L F^R(s) = -\Gamma^p(s) \dot{A}_{t-s},
\end{align}
\end{subequations}
where we defined the memory kernel due to the $P_p$ projection as
\begin{equation}
\Gamma^p(s) = \frac{\langle F^R(0), F^R(s) \rangle}{\langle \dot{A}_0^2\rangle}.
\label{eq_Gammap}
\end{equation}
The memory friction due to the $P_x$ projection can,
using eq. \eqref{eq_propagation_cond_avrg},
be written as a conditional average 
\begin{equation}
\label{eq_Px_memory}
e^{(t-s)L}P_x L F^R(s) 
%=e^{(t-s)L}  \langle L F^R(s)\rangle_{A_{0}} 
= \langle L F^R(s)\rangle_{A_{t-s}} \equiv \Gamma^x (A_{t-s}, s),
\end{equation}
which, using  the relation in eq.~\eqref{eq_cond_avrg_rel}, can be rewritten as 
\begin{equation}
\label{eq_Gammax}
\Gamma^x(A_t,s) = \frac{\mathrm{d}}{\mathrm{d}A_t} D(A_t,s)-\beta D(A_t,s)\frac{\mathrm{d}}{\mathrm{d}A_t} U_\mathrm{PMF}(A_t).
\end{equation}
Here,  we introduced the conditional correlation function between the time derivative of the observable at the initial time,
 $\dot A_0$, and the random force $F^R(s)$
\begin{equation}
\label{eq_D}
D(A_t, s) = \langle \dot{A}_0, F^R(s) \rangle_{A_t}.
\end{equation}
%Note that $D(A_t, s)$ is a function of two variables only. 
%

With the definition of the hybrid projection operator $P_H$ in eq. \eqref{eq_projection} and 
the results in eq.~\eqref{eq_genU}, eq.~\eqref{eq_Pp_memory} and eq.~\eqref{eq_Px_memory}, the general GLE in
eq.~\eqref{eq_GLE_1} takes the specific form 
\begin{align}
\label{eq_main_result1}
&\ddot{A}_t = -\frac{1}{M(A_t)}\frac{\mathrm{d}}{\mathrm{d}A_t}\left( U_\mathrm{PMF}(A_t) + k_BT\ln M(A_t)\right)\nonumber\\
& - \int_0^t\mathrm{d}s\,\Gamma^p(s)\dot{A}_{t-s} + \int_0^t\mathrm{d}s\,\Gamma^x(A_{t-s}, s)  + F^R(t),
\end{align}
which is the exact GLE that follows from our hybrid projection scheme and constitutes a main result of our paper.
 A few comments are in order: 
i) The PMF $U_\mathrm{PMF}(A_t)$ appears explicitly in the equation of motion, similar to  the Zwanzig projection scheme.
ii)  An inhomogeneous effective mass  $M(A_t)$ gives rise  to a drift term.
 If  $M(A_t)$  is constant, i.e., if the variance of  $\dot{A}_t$ is independent of  $A_t$, see 
  eq. \eqref{eq_gen_M}, this drift term vanishes.
 For an observable $A_t$ that is a  linear combination of positions, it follows directly that the effective mass is constant \cite{glatzel_interplay_2021}.
Even  for certain non-linear observables, such as distances in position space, 
it can be shown that the generalized mass is constant, as  demonstrated  in appendix~\ref{sec_App_Mass}. 
On the other hand, for angles, which are three-body terms, the effective mass will in general depend on $A_t$, as demonstrated
for the dihedral angle of butane in section \ref{sec_Apl_dihedral}.
iii) The memory kernel $\Gamma^p(s)$  is determined via the unconditional average over the random-force correlations in 
eq. \eqref{eq_Gammap}, similarly to the Mori projection, and  therefore only depends on time. It thus
describes the linear friction contribution.
iv) The  memory friction function $\Gamma^x(A_{t-s},s)$ is a general function of  the observable $A_{t-s}$, 
it therefore accounts for non-linear friction contributions. 
According to eq.~\eqref{eq_Gammax}, this contribution disappears if the
conditional correlation  function between the random force and the time derivative of the observable, $D(A_t, s)$,  as defined
in eq. \eqref{eq_D}, vanishes.
This constitutes the exact condition for which the approximate  GLE in eq. \eqref{eq_GLE_adhoc} is valid. 
v) The first moment of the random force vanishes,  $\langle F^R(t) \rangle = 0$,  as follows from the relation eq.~\eqref{eq_avereage_orthogonal_dyn}.
The second moment is determined by the memory kernel $\Gamma^p(s)$ via eq. \eqref{eq_Gammap}. 
Higher cumulants do  not necessarily vanish but are not expected to play a significant role since non-linear effects are
already accounted for by the PMF $U_\mathrm{PMF}(A_t)$.  Indeed, in section \ref{sec_Apl_dihedral} we demonstrate for the explicit example of the butane dihedral angle that the random-force distribution  exhibits finite but moderate non-Gaussian contributions. 

%Note that, spatial non-homogeneities would have an impact on the conditional velocity distribition $\mathbb{P}(\ve p_k|\ve r_k)$, i.e., the probability to observe a momentum $\ve p_k$ given a position $\ve r_k$. If the medium is homogeneous, the conditional velocity distribution will be the same for all positions $\ve r_k$ and thus, $\langle \hat{\ve p}_k, F_n(s)\rangle_{\ve r_k=\ve a} = \langle \hat{\ve p}_k, F_n(s)\rangle = 0$.

The multi-dimensional generalization of eq.~\eqref{eq_main_result1}, i.e., the case in which the observable 
 is a vector $\mathbf{A}(\mathbf{R}_t) = (A_1(\mathbf{R}_{t}), A_2(\mathbf{R}_{t}),\dots, A_n(\mathbf{R}_{t}))$,
  is derived  in appendix~\ref{sec_App_multidim}. 

\section{Numerical Scheme For Extracting Random Forces From Trajectories\label{sec_proj_corr_func}}

In the absence of a potential and in the absence of non-linear friction, 
Carof et al. presented iterative algorithms to compute the random force trajectory and the linear friction kernel from a trajectory
of the reaction coordinate
 \cite{carof_two_2014, lesnicki_molecular_2016}. 
%Thereby, they were able to compute the random force $F^R(\omega_0,t)$ without computing the memory kernel beforehand.
%Their method allowed them to study the statistical properties of the random force and identify deviations from Gaussian statistics. 
Their derivations explicitly use the Mori projection, so the results are only valid for the Mori GLE in eq.~\eqref{eq_mori_GLE}. 
%An extension to the ad-hoc GLE was given in later works \cite{carof_coarse_2014, lesnicki_molecular_2016} by a wrong identification of external forces with a potential of mean force, neglecting hereby the average force exerted from internal degrees of freedom. 

We now introduce a method to compute the random force trajectory $ F^R(\omega_0,t)$ and from that the memory kernel $\Gamma^p(t)$ and the  non-linear memory function  $\Gamma^x(A_s,t-s)$  
as defined by our  GLE,  eq.~\eqref{eq_main_result1},   from a given trajectory of an arbitrary observable. 
 For this, let us consider the projected propagator $e^{tQ_H L}$ based on our hybrid projection scheme eq.~\eqref{eq_projection}.
 From the Dyson decomposition in eq.~\eqref{eq_dyson_decomposition}, we obtain by rearranging 
\begin{align}
\label{eq_projected_1}
e^{tQ_HL} &= e^{t L} - \int_0^t\mathrm{d}s\,e^{(t-s)L}P_HLe^{sQ_HL}.
\end{align}
Applying eq.~\eqref{eq_projected_1} on the  initial random force $F^R(\omega_{0},0)$ and using 
eq. \eqref{eq_F_operator} and
the memory functions $\Gamma^p(t)$ and $\Gamma^x(A,t)$ defined in eq.~\eqref{eq_Pp_memory} and eq.~\eqref{eq_Px_memory}, respectively, we find
\begin{align}
\label{eq_projected_FR_t}
F^R(\omega_{0},t) &= e^{tL}F^R(\omega_0,0) + \int_0^t\mathrm{d}s\,\Gamma^p(s)e^{(t-s)L}\dot{A}_0\nonumber\\
& \quad - \int_0^t\mathrm{d}s\,e^{(t-s)L}\Gamma^x(A_0, s).
\end{align}
Now, we consider eq.~\eqref{eq_projected_1} at time $t+\Delta t$
\begin{align}
e^{(t+\Delta t)Q_H L} &= e^{t L}e^{\Delta t L}\nonumber\\
& \quad  - \int_0^{t+\Delta t}\mathrm{d}s\,e^{(t-s)L}e^{\Delta tL}P_H Le^{sQ_H L}.
\end{align}
Buy splitting up the integral on the r.h.s. into two parts, we obtain
\begin{align}
\label{eq_projected_12}
e^{(t+\Delta t)Q_H L} &= e^{t L}e^{\Delta t L} - \int_0^{t}\mathrm{d}s\,e^{(t-s)L}e^{\Delta tL}P_H Le^{sQ_H L}\nonumber\\
& \quad - \int_0^{\Delta t}\mathrm{d}s\,e^{(\Delta t-s)L}P_H Le^{(t+s)Q_H L},
\end{align}
where we used the substitution $s\to s-t$ in the second integral. Acting with the operator in eq.~\eqref{eq_projected_12} on the
initial  random force $F^R(\omega_0, 0)$ and using eq.~\eqref{eq_propagator_prob} gives
\begin{align}
\label{eq_projected_2}
&F^R(\omega_{0},t+ \Delta t) = e^{tL} F^R(\omega_{\Delta t},0)\nonumber\\ 
&\quad + \int_0^t \mathrm{d}s\, \Gamma^p(s) e^{(t-s)L} \dot{A}_{\Delta t} - \int_0^t\mathrm{d}s\,e^{(t-s)L} \Gamma^x(A_{\Delta t}, s)\nonumber\\
&\quad - \int_0^{\Delta t}\mathrm{d}s\,e^{(\Delta t-s)L}P_H L F^R(\omega_0,t+s).
\end{align}
Comparing eq.~\eqref{eq_projected_2} with eq.~\eqref{eq_projected_FR_t}, we see that the first three terms on the r.h.s. of eq.~\eqref{eq_projected_2} are equal to $F^R(\omega_{\Delta t},t)$. Hence, we find 
\begin{align}
\label{eq_projected_4}
F^R(\omega_{0},t+\Delta t) &= F^R(\omega_{\Delta t}, t) +\int_0^{\Delta t}\mathrm{d}s\,\Gamma^p(t+s) \dot{A}_{\Delta t-s}\nonumber\\
&\quad - \int_0^{\Delta t}\mathrm{d}s\,\Gamma^x(A_{\Delta t-s},t+s).
%&\Gamma^p_{nl}(t+s) = \frac{\langle F^R_n(\hat{\omega}_{0},0), F^R_l(\hat{\omega}_{0},t+s)\rangle}{\langle \hat p_{l,0}^2\rangle},\\
%&\Gamma^x_n(\ve r_{\Delta t-s}, t+s) = \left[\nabla_r\cdot\mathbf{D}_n(\mathbf{r},t+s)-\beta\, \mathbf{D}_n(\mathbf{r},t+s)\cdot \nabla_r U_\mathrm{PMF}(\mathbf{r})\right]_{\ve r = \ve r_{\Delta t - s}},\\
%&\mathbf{D}_n(\mathbf{r},t) = \frac{\langle \delta(\hat{\mathbf{r}}_0-\mathbf{r})\frac{\hat{\mathbf{p}}_0}{m}, F^R_n(\hat\omega_{0},t)\rangle}{\langle \delta(\hat{\mathbf{r}}_0-\mathbf{r})\rangle}.
\end{align}
For given  trajectories  $A_t, \dot{A}_t$ and given random  force $F^R(\omega_{\Delta t},t)$ as a function of the phase space configuration 
$\omega_{\Delta t}$, eq.~\eqref{eq_projected_4} gives the random force $F^R(\omega_0,t+\Delta t)$ one time step $\Delta t$ later 
as a function of the phase space configuration $\omega_0$ one time step $\Delta t$  before. 
%\begin{align}
%\label{eq_projected_3}
%e^{(t+\Delta t)QL}\ve F^R(\omega_{t_0},0) &= e^{tQL}e^{\Delta t L}\ve F^R(\omega_{t_0},0) - %\int_0^{\Delta t}\mathrm{d}s\,e^{(\Delta t-s)L}PLe^{(t+s)QL}\ve F^R(\omega_{t_0},0).
%\end{align}
To obtain an iterative scheme for the random force, 
eq.~\eqref{eq_projected_4} is  discretized in time and $A$-space. 
For this, we use the left rectangular rule to discretize the time integrals. 
The random fore is discretized as $F^R(\omega_{t^\prime}, t)=F^R(\omega_{i\Delta t}, j \Delta t) \equiv F^R(i,j)$. 
 The $A$-space is divided into  $N_A$ bins with width $\Delta A$,  
the bin intervals are labeled by $I_\alpha =\left[\alpha \Delta A, (\alpha + 1)\Delta A\right]$ with $\alpha=1,2,\dots,N_A$.
The discretized versions of eqs.  \eqref{eq_Gammap},  \eqref{eq_Gammax},   \eqref{eq_D} and ~\eqref{eq_projected_4}  read
\begin{widetext}
\begin{subequations}
\label{eq_projected_forward}
\begin{align}
&F^R(i,j+1) = F^R(i+1, j) +\Delta t\, \Gamma^p(j) \dot{A}_{i+1} - \Delta t\, \Gamma^x(A_{i+1},j) + \mathcal{O}(\Delta t^2),\label{eq_projected_forward_a}\\
&\Gamma^p(j) = \frac{\sum_{i=0}^{N_\mathrm{traj}-j-1} F^R(i,0) F^R(i,j)}{\sum_{i=0}^{N_\mathrm{traj}-j-1}\dot{A}_{i}^2},\label{eq_projected_forward_b}\\
&\Gamma^x(A_{i+1}, j) = \left[\frac{D(\alpha+1,j)-D(\alpha-1,j)}{2\Delta A}-\beta\, D(\alpha,j) \frac{U_\mathrm{PMF}(\alpha+1)-U_\mathrm{PMF}(\alpha-1)}{2\Delta A}\right]_{A_{i+1}\in I_\alpha},\label{eq_projected_forward_c}\\
&D(\alpha,j) = \frac{\sum_{\substack{i\leq N_\mathrm{traj}-j-1\\ A_i\in I_\alpha}}\dot{A}_i F^R(i,j)}{\sum_{\substack{i\leq N_\mathrm{traj}-j-1\\ A_i\in I_\alpha}} 1}.\label{eq_projected_forward_d}
\end{align}
\end{subequations}
\end{widetext}
If the observable $A_t$ has at time $t=i\Delta t$ a value in the interval $I_\alpha$, we write $A_i\in I_\alpha$;
 $\sum_{A_i\in I_\alpha}$ denotes the sum over all times $i$ for which  $A_i$ is in the interval $I_\alpha$,
 which is used to compute  conditional averages in eq.~\eqref{eq_projected_forward}.
%The propagator $e^{\Delta t L}$ acts on the phase space variables, i.e., $e^{\Delta t L}\ve F^R(\omega_{t_0},0) = \ve F^R(\omega_{t_0+\Delta t},0)$ while $e^{tQL}$ propagates the function explicitly in time, i.e., $e^{tQL}\ve F^R(\omega_{t_0+\Delta t},0) = \ve F^R(\omega_{t_0+\Delta t},t)$. Following these rules, we get as a result
$N_\mathrm{traj}$ denotes the total  length of the  $A_t$ trajectory used. 
The sums run from $i=0$ to $N_\mathrm{traj}-j-1$, because  for given $j$, the iterative scheme has only
determined  the random force at times up to $N_\mathrm{traj}-j-1$,
as  follows from eq.~\eqref{eq_projected_forward_a}. 
The sums in the denominator extend over the same interval as in the numerator in order to increase the  numerical  stability  
 \cite{carof_two_2014, lesnicki_molecular_2016}. 
The derivatives in $A$-space in eq.~\eqref{eq_projected_forward_c} are computed using central differences.
The iterative scheme in eq.~\eqref{eq_projected_forward} works as follows: 
First, note from eq.~\eqref{eq_main_result1}  that 
$F^R(i,0)=\ddot{A}_i + (1/M(A_i))\mathrm{d}/\mathrm{d}A_i\left[U_\mathrm{PMF}(A_i)+k_BT\ln M(A_i)\right]$, i.e., the random
force at time $t=0$  equals  the acceleration plus the force from the effective  potential for all possible
initial times $i\Delta t$ for $i=0,1,2,\dots, N_\mathrm{traj}-1$. 
This, together with  $\dot{A}_i$, can be obtained directly from a given trajectory of the observable $A$. 
Then, $F^R(i,0),A_i$, and $\dot{A}_i$ are inserted into eq.~\eqref{eq_projected_forward} to compute $F^R(i,1)$ 
for $i=0,1,2,\dots,N_\mathrm{traj}-2$. $F^R(i,1)$ is then used to compute $F^R(i,2)$ for $i=0,1,2,\dots,N_\mathrm{traj}-3$ and so forth. 
While computing $F^R(i,j)$, the memory friction functions $\Gamma^p(i)$ and $\Gamma^x(A_i,i-j)$ are computed simultaneously. 
If our only goal is to compute the  memory friction functions, 
 we can stop the computation of $F^R(i,j)$ as soon as the memory functions have dropped to zero. 
 As an example, if the memory functions decay to zero after $N_\mathrm{mem}$ time steps, we can abort the computation of 
 the random force at $F^R(i,N_\mathrm{mem})$. At that point, we generated $N_\mathrm{traj}-N_\mathrm{mem}-1$ distinct random-force trajectories of length $N_\mathrm{mem}$
 each. Since the memory functions are computed simultaneously, the generated random-force trajectories only need to be stored 
 if one is interested in the random-force statistics, in which case one could extend the length of the random-force trajectories. 
 In appendix~\ref{sec_App_trapz}, we present an alternative discretization in time for eq.~\eqref{eq_projected_4}
 using the trapezoidal rule.

\section{Applications}

\begin{figure*}[t]
\includegraphics[width=0.9\textwidth]{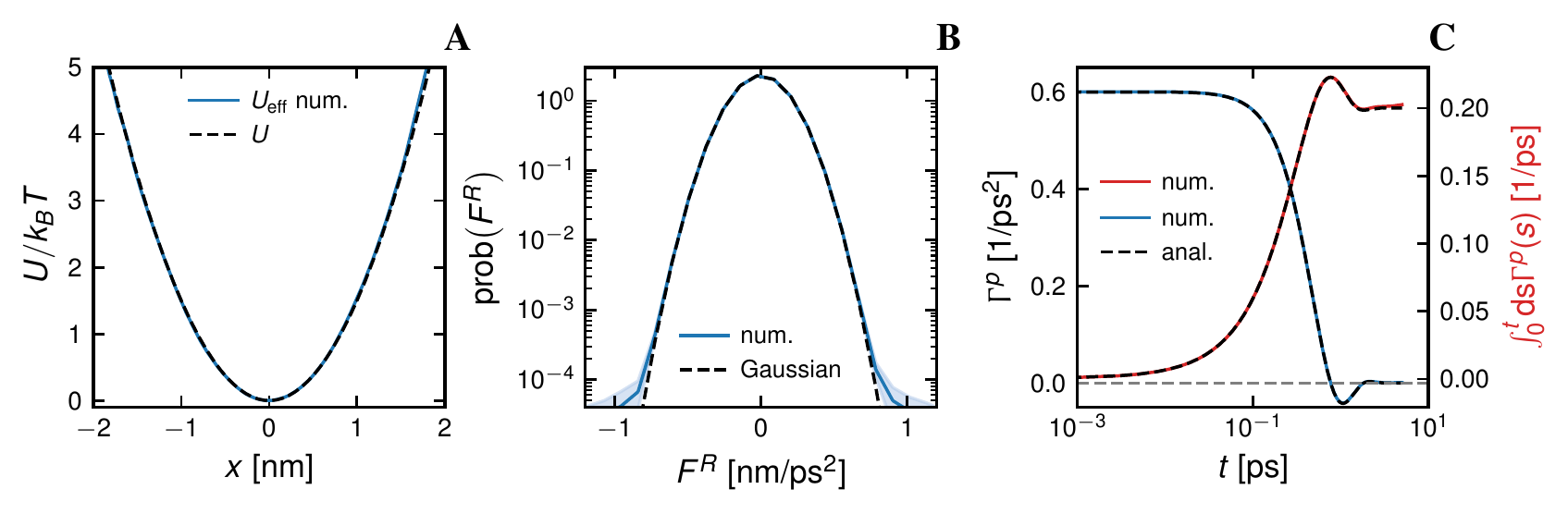}
\caption{\label{fig_1}Test of  the numerical extraction scheme  in eq.~\eqref{eq_projected_forward} using the harmonic Hamiltonian model 
defined in eq.~\eqref{eq_Zwanzig_model_nonlocal}. 
In \textbf{A}, 
we compare the input potential $U(x)=kx^2/2$ (broken line) with the numerically obtained effective potential $U_\mathrm{eff}(x)$ defined in 
eq. \eqref{eq_gen_U1} (solid blue line).
In  \textbf{B}, we confirm that the numerical extraction of the random force leads to the expected 
Gaussian distribution with zero mean and  standard deviation $\sqrt{\langle p_0^2\rangle\,\Gamma^p(0)}$. The shaded area in blue highlights the numerical error.
In \textbf{C}, we check that  the  analytical memory function $\Gamma^p(t) $
 given in  eq. \eqref{eq_Gammap_stochastic} and its running integral are accurately reproduced by the 
 numerical results from the extraction scheme.}
\end{figure*}

We test our numerical algorithm in eq.~\eqref{eq_projected_forward} on three different systems: an exactly solvable harmonic Hamiltonian model which leads to a GLE without spatial dependencies in the memory friction  term, the non-linear Hamiltonian version of the 
Zwanzig model \cite{zwanzig_nonlinear_1973}, where spatial dependencies in the memory friction  term are present, and finally, 
we discuss results obtained for the dihedral angle dynamics of a butane molecule in water from MD simulations.

\subsection{Harmonic Hamiltonian Model\label{sec_Apl_harmonic}}

\begin{figure}
\includegraphics[width=0.7\linewidth]{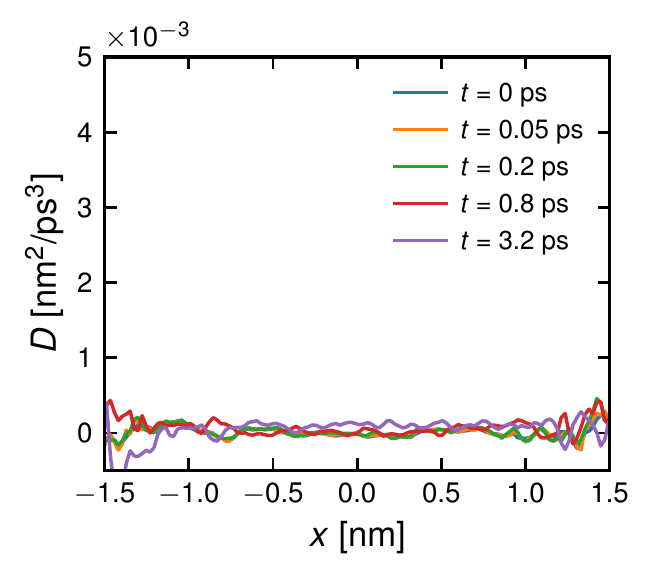}
\caption{\label{fig_2} Conditional velocity-random force correlation function $D(x,t)$ defined in eq.~\eqref{eq_D} for the harmonic model defined in eq.~\eqref{eq_Zwanzig_model_nonlocal}.}
\end{figure}

\begin{figure*}[t]
\includegraphics[width=0.9\textwidth]{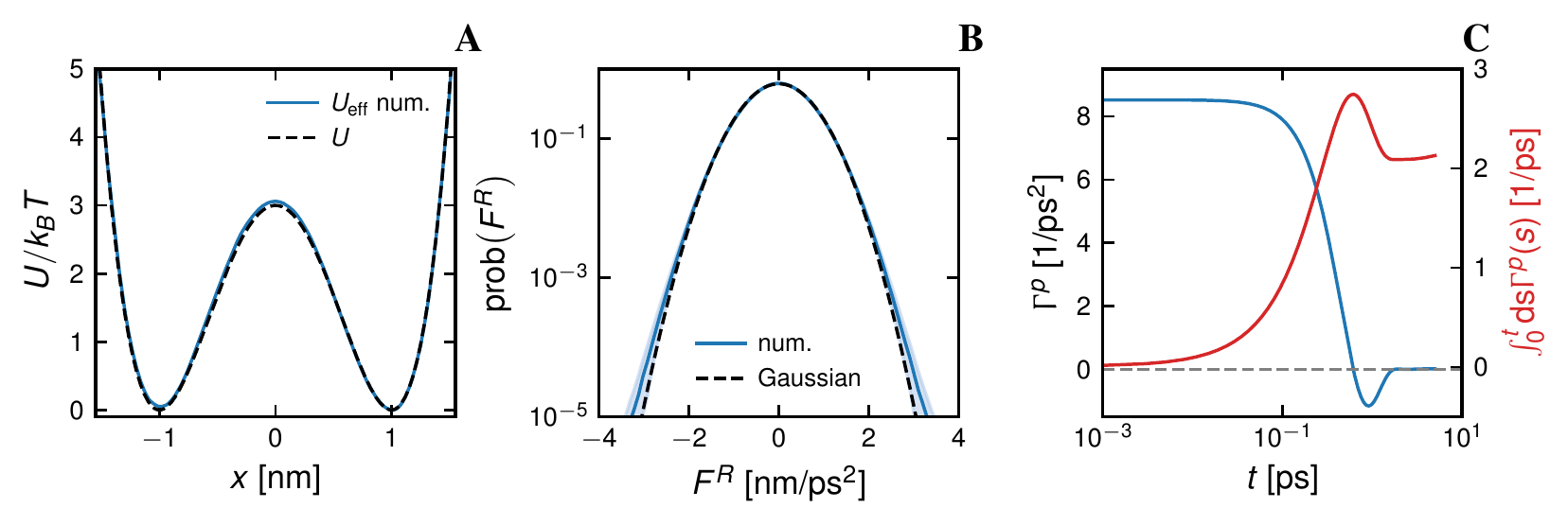}
\caption{\label{fig_3}Test of  the numerical algorithm in eq.~\eqref{eq_projected_forward} for  the non-linear  Zwanzig model  defined in eq.~\eqref{eq_Zwanzig_model_local}  using a double-well potential $U(x)=U_0(x^2-1)^2$. 
\textbf{A} Comparison of the input 
 potential $U(x)$ (broken line) with the numerically obtained effective potential $U_\mathrm{eff}(x)$ defined in 
eq. \eqref{eq_gen_U1} (solid blue line). The two potentials coincide, which means that the effective mass is constant,
as expected on analytic grounds.
\textbf{B}
The numerically extracted random force distribution (blue solid  line)  is well described by a Gaussian with vanishing mean 
and  standard deviation  $\sqrt{\langle \dot{x}^2_0\rangle\, \Gamma^p(0)}$ (broken black line). 
The shaded area in blue highlights the numerical error.
In \textbf{C}, we show the numerically extracted  memory function $\Gamma^p(t) $ and its running integral.}
\end{figure*}

The exactly solvable harmonic model is defined by the Hamiltonian
\begin{align}
\label{eq_Zwanzig_model_nonlocal}
H(x,p,\lbrace q_n,v_n\rbrace ) &= \frac{p^2}{2 m} +\sum_{n=1}^N\frac{v_n^2}{2 m_n} + U(x)\nonumber\\
& \quad + \sum_{n=1}^N\frac{k_n}{2}(x-q_n)^2.
\end{align}
The relevant coordinates are the one-dimensional position $x$ and momentum $p$
which are coupled to the auxiliary particle positions $q_n$ and momenta $v_n$.
 If we choose the potential $U(x)$ to be a harmonic potential, i.e., 
 $U(x) = k x^2/2$, we can use our hybrid projection formalism to exactly derive the GLE. 
 For this we compute the random force $F^R(t)$ defined  in eq.~\eqref{eq_F_operator} by 
 an operator expansion to all orders
\begin{align}
F^R(t) = e^{tQ_HL}Q_HLp_0 = \sum_n^\infty\frac{t^n}{n!}(Q_HL)^n\, Q_HLp_0,
\end{align}
as shown in appendix~\ref{sec_App_POM}. Once $F^R(t)$ is computed, the memory functions $\Gamma^p(t)$ and $\Gamma^x(x,t)$ are obtained from eq.~\eqref{eq_Gammap} and eq.~\eqref{eq_Gammax}, respectively. 
In appendix~\ref{sec_App_ZwanzigGLE} we show how to alternatively obtain a GLE without projection,
namely by solving the equations of motion for the $q_n$ variables  and inserting the result back into the equation of motion for $x$,
which works for general potential $U(x)$.
%This results in a GLE of the approximate form in eq.~\eqref{eq_GLE_adhoc}. 
The GLE's obtained from the projection formalism and the exact solution  agree with one another and 
take the form of the approximate GLE in eq.~\eqref{eq_GLE_adhoc},
\begin{align}
\dot{p}_t&=-U^\prime_\mathrm{PMF}(x_t)-\int_0^t\mathrm{d}s\,\Gamma^p(t-s) p_s+F^R(t).
\end{align} 
The memory friction kernel is given by
\begin{align}
\label{eq_Gammap_FT}
\Gamma^p(t) = \frac{1}{m}\sum_{n=1}^N k_n \cos(\mu_n t)
\end{align}
with $\mu_n = \sqrt{k_n/m_n}$. 
We note that the spatially dependent memory friction term $\Gamma^x(x,t)$ vanishes, as shown in 
 appendix~\ref{sec_App_POM} and \ref{sec_App_ZwanzigGLE}. 
 
% Further, one finds that the distribution of the random force $F^R(t)$ at all times $t$ is a Gaussian with vanishing mean and standard deviation $\sqrt{\langle v^2\rangle \Gamma^p(0)}$ which can also be checked numerically using eq.~\eqref{eq_projected_forward} in combination with a trajectory $x(t)$. 
 
In order to test our numerical scheme in eq.~\eqref{eq_projected_forward}, we need to generate trajectories of $x_t$. 
To do this in a numerically efficient fashion, 
we identify eq.~\eqref{eq_Gammap_FT} as the Fourier series of an even function with Fourier coefficients $k_n$. 
In the limit of $N\to\infty$ and for  a continuous frequency dependency, i.e., $k_n \to k(\mu)\mathrm{d}\mu/2\pi$, we can choose
the exponential-oscillating memory kernel 
\begin{align}
\label{eq_Gammap_stochastic}
\Gamma^p(t) = \frac{K}{m}\, e^{-\frac{|t|}{2\tau_\Gamma}}\left(\cos\left(\frac{\nu}{2 \tau_\Gamma} t\right)+\frac{1}{\nu}\sin\left(\frac{\nu}{2 \tau_\Gamma} |t|\right)\right).
\end{align} 
This maps the Hamiltonian system in eq.~\eqref{eq_Zwanzig_model_nonlocal} onto the stochastic system of  two
linearly coupled Langevin equations
\begin{subequations}
\label{eq_stochastic_1}
\begin{align}
m \ddot{x}_t &= -k x_t-K(x_t-y_t),\\
m_y \ddot{y}_t &= -K(y_t-x_t)-\gamma \dot{y}_t + \sqrt{2 k_BT\gamma}\eta(t),\\
\langle \eta(t)\rangle &=0, \qquad \langle \eta(t),\eta(0)\rangle = \delta(t),
\end{align}
\end{subequations}
where  $\eta(t)$ in eq.~\eqref{eq_stochastic_1} is a white noise variable,
as derived  in appendix~\ref{sec_App_StochasticZwanzig}. The parameters in eq.~\eqref{eq_Gammap_stochastic} and eq.~\eqref{eq_stochastic_1} are related by $\tau_\Gamma = m_y/\gamma$ and $\nu^2 = 4 m_y K /\gamma^2-1$. The scalar
$y$ variable in eq.~\eqref{eq_stochastic_1} is the stochastic representation of the Hamiltonian environment produced  by the $q_n$  variables in eq.~\eqref{eq_Zwanzig_model_nonlocal}. Using eq.~\eqref{eq_stochastic_1}, we numerically
generate trajectories $x_t$
for a system with thermal energy $k_BT = 2.5\,$kJ/mol, which corresponds to $T=300\,$K  \cite{pronk2013gromacs}. 
The other parameters are chosen to be $m=50\,$u, $m_y=2\,$u, $K=30\,$kJ/mol/nm, $k=7.5\,$kJ/mol/nm, $\gamma=10\,$u/ps and a time step of $\mathrm{d}t=0.001\,$ps. The simulation time is 100\,ns. The results shown in fig.~\ref{fig_1}B, fig.~\ref{fig_1}C and fig.~\ref{fig_2} are obtained by averaging over the results of 100  independent trajectories.

In fig.~\ref{fig_1} we compare analytical results with   results  derived from the numerically generated trajectories using the scheme in eq.~\eqref{eq_projected_forward}, where the $x$-space is discretized using $N_A=200$ bins of equal length. 
In fig.~\ref{fig_1}A we compare the input potential $U(x)=kx^2/2$ (broken line) with the numerically obtained effective potential 
$U_\mathrm{eff}(x)$ defined in 
eq. \eqref{eq_gen_U1} (solid blue line), both potentials are shifted so that they are zero at $x=0$. The agreement is perfect, 
which in particular means that the effective mass $M(x)$ defined in eq. \ref{eq_gen_M} is a constant, as expected. 
In fig.~\ref{fig_1}B we compare  the analytical and the numerically determined  random force distribution, which demonstrates that indeed
non-Gaussian contributions are absent. In fig.~\ref{fig_1}C we compare the analytic memory kernel  $\Gamma^p(t)$
in eq. \eqref{eq_Gammap_stochastic} with the one extracted from the simulation trajectory and again obtain perfect agreement.
This all shows that the numerical extraction scheme works perfectly on fluctuating trajectories. 
 In fig.~\ref{fig_2} we show the numerical result for the function $D(x,t)$ defined by eq.~\eqref{eq_Px_memory} for different times. 
  As predicted in  appendix~\ref{sec_App_POM},  $D(x,t)$ vanishes for all times.

\subsection{The Non-Linear Zwanzig Model\label{sec_Apl_nonlinear}}

\begin{figure}
\includegraphics[width=\linewidth]{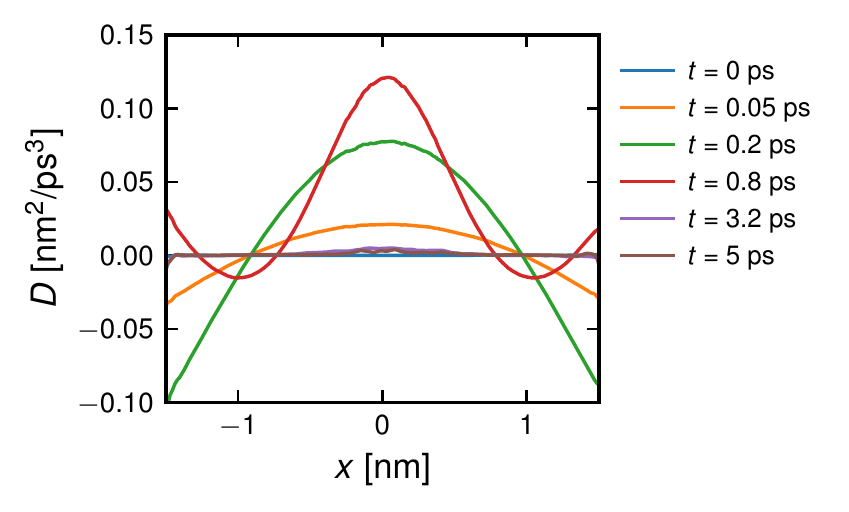}
\caption{\label{fig_4} 
Conditional velocity-random force correlation function $D(x,t)$ defined in eq.~\eqref{eq_D} for the non-linear Zwanzig model defined in eq.~\eqref{eq_Zwanzig_model_local} as a function of the position $x$ for different times.}
\end{figure}

\begin{figure*}
\includegraphics[width=0.9\textwidth]{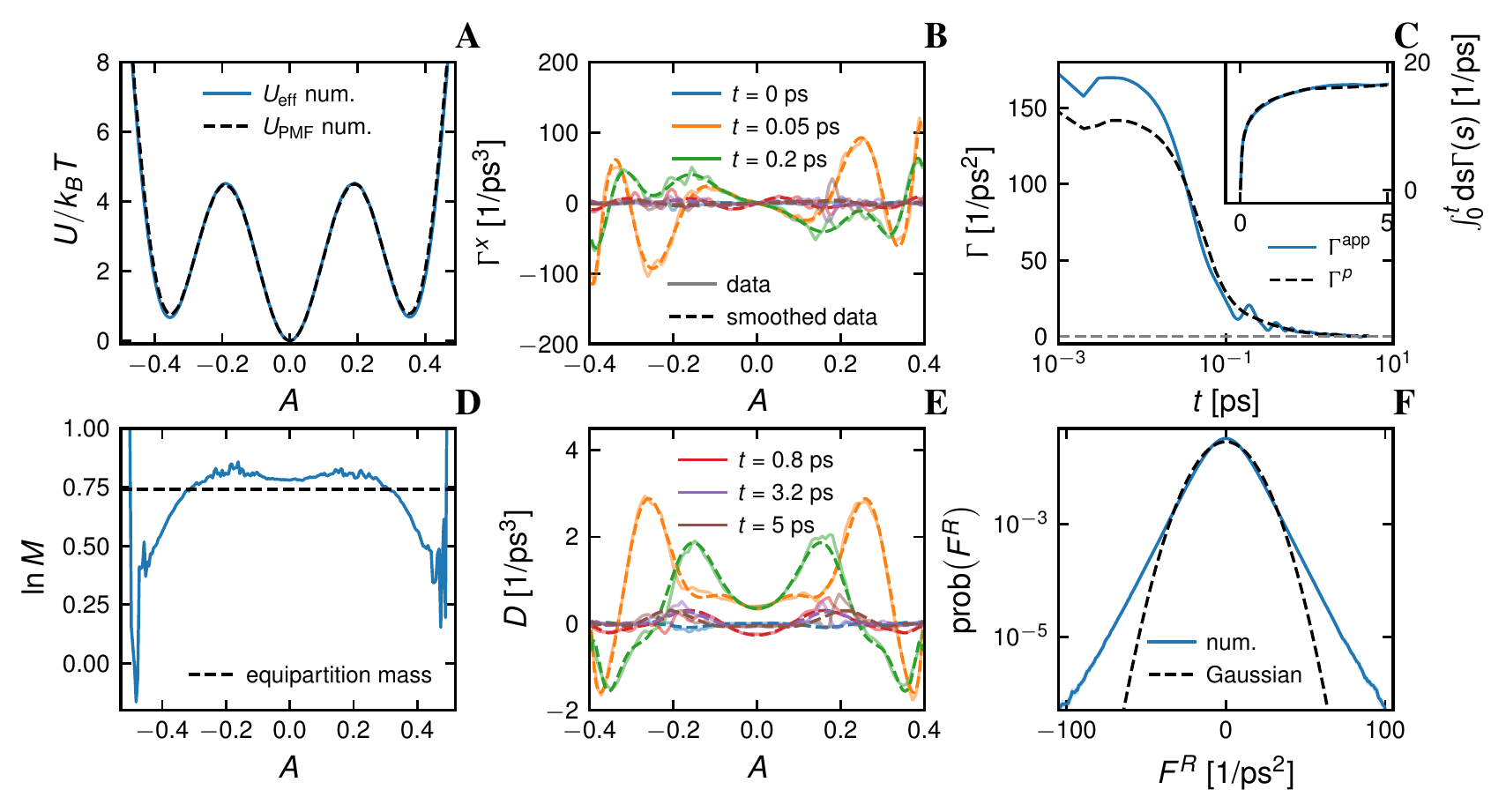}
\caption{\label{fig_5}
Extraction of GLE parameters for MD data of the dihedral angle dynamics of butane in water. 
Here, $A=\phi/(\phi_\mathrm{max}-\phi_\mathrm{min}) \in (-0.5,0.5)$ denotes the rescaled dihedral angle with $\phi$ being the dihedral angle and $\phi_\mathrm{max}$, $\phi_\mathrm{min}$ being the largest and smallest value of $\phi$ along the trajectory, respectively. 
\textbf{A}  
A small  deviation between the effective potential $U_\mathrm{eff}(A)$ defined in  eq. \eqref{eq_gen_U1}  (blue solid line) and the numerically computed potential of mean force  $U_\mathrm{PMF}(A)$ (broken line) is observed, which is explained by the dependence of the effective mass $M(A)$ on the dihedral angle, as  shown in  D. 
\textbf{B}
 Non-linear friction function $\Gamma^x(A,t)$ defined by  eq.~\eqref{eq_Gammax} 
for different times. Solid lines correspond to the original data, the dashed lines show smoothed data.
\textbf{C}
Comparison of the memory function $\Gamma^p(t)$ defined within the hybrid GLE eq. \eqref{eq_main_result1}
and the memory function $\Gamma^\mathrm{app}(t)$ defined within the approximate
 GLE in eq.~\eqref{eq_GLE_adhoc}. The inset shows the respective running integrals. 
\textbf{D} Logarithm of the effective mass $M(A)$ as defined by eq. \eqref{eq_gen_M}.  
\textbf{E}
Correlation function $D(A,t)$ defined in eq.~\eqref{eq_D} as a function of  $A$ for different times. Solid lines correspond to the original data, the dashed lines show smoothed data.
\textbf{F} The extracted  random force distribution (blue line) shows pronounced deviations from a Gaussian with vanishing mean and standard deviation of $\sqrt{\langle \dot{A}^2_0\rangle \Gamma^p(0)}$ (broken line). 
The shaded area in blue highlights the numerical error and is of the order of the line width.}
\end{figure*}

As the second exactly solvable model we consider 
the Hamiltonian version of the non-linear Zwanzig model \cite{zwanzig_nonlinear_1973}, for which non-linear friction effects are present
and therefore the 
 approximate  GLE in eq.~\eqref{eq_GLE_adhoc} is not valid anymore.
This model is defined by the Hamiltonian 
\begin{align}
\label{eq_Zwanzig_model_local}
H(x,p,\lbrace q_n,v_n\rbrace ) &= \frac{p^2}{2 m} +\sum_{n=1}^N\frac{v_n^2}{2 m_n} + U(x)\nonumber\\
& \quad  + \sum_{n=1}^N\frac{k_n}{2}(\alpha(x)-q_n)^2.
\end{align}
In eq.~\eqref{eq_Zwanzig_model_local}, a generally non-linear function $\alpha(x)$ determines the coupling between 
the relevant variable $x$ and the auxiliary variables $q_n$. Note that for $\alpha(x)=x$, we obtain back the harmonic model defined 
in eq.  \eqref{eq_Zwanzig_model_nonlocal}.
%The model is discussed in more detail in appendix~\ref{sec_App_ZwanzigGLE}. 
The GLE that follows from the
Hamiltonian system in eq.~\eqref{eq_Zwanzig_model_local} can not be calculated in closed form using our hybrid projection scheme
for general $\alpha(x)$,
we therefore cannot derive the exact form of $\Gamma^p(t)$. 
On the other hand, by solving the equations of motion for the $q_n$  variables and inserting the result into the equation for $x$, 
one finds a GLE of the form
\begin{subequations}
\begin{align}
\label{eq_nonlinear_GLE}
m \ddot{x}_t &= -U^\prime(x_t) - \int_0^t\mathrm{d}s\,\Gamma\left[t-s,x_t,x_s\right]\dot{x}_s + F^R_Z(t, x_t),
\end{align}
with a (for $\alpha(x)\neq x$) non-linear  memory friction function
\begin{align}
\Gamma(t-s,x_t,x_s) & = \alpha^\prime(x_t) \alpha^\prime(x_s)\sum_{n=1}^N k_n \cos(\mu_n(t-s))
\label{eq_nonlinear_mem}.
\end{align}
\end{subequations}
Actually, the form of the memory kernel $\Gamma(t-s,x_t,x_s)$ in eq. \eqref{eq_nonlinear_mem}, and  in particular its dependence on 
the trajectory $x_t$, is not compatible with the form of the memory function $\Gamma^x(x_{t-s},s)$ or, equivalently, 
$\Gamma^x(x_{s},t-s)$, in eq.  \eqref{eq_main_result1}.
In fact, 
in appendix~\ref{sec_App_Mapping_GLEs} we demonstrate that  the GLEs given in eq. \eqref{eq_nonlinear_mem}
and in eq.  \eqref{eq_main_result1} are equivalent in the sense that they produce, for identical initial conditions, 
identical trajectories $x_t$. This of course is expected, since they follow via  exact derivations from the same Hamiltonian.
This finding is similar to the fact that the Mori and Zwanzig GLEs are, in the absence of  
approximations, also equivalent and shows that even GLEs with  identical PMFs and different friction memory and
random force terms can be equivalent. 
For $\alpha(x)\neq x$, it is therefore interesting to extract the non-linear friction term  $\Gamma^x(x_{t-s},s)$,
as  defined by our 
GLE in eq. \eqref{eq_nonlinear_mem}, from simulation trajectories of $x_t$. 
 
Similar to our approach to obtain eq.~\eqref{eq_stochastic_1} for $N\to\infty$, we  exploit the structure of eq.~\eqref{eq_nonlinear_mem},
which is equivalent to a Fourier decomposition in the time domain, 
to map the Hamiltonian system in eq.~\eqref{eq_Zwanzig_model_local} onto a system of non-linearly 
coupled Langevin equations given by (see appendix~\ref{sec_App_StochasticZwanzig})
\begin{subequations}
\label{eq_stochastic_2}
\begin{align}
m \ddot{x}_t &= -U^\prime(x_t)-K\alpha^\prime(x_t)(\alpha(x_t)-y_t),\\
m_y \ddot{y}_t &= -K(y_t-\alpha(x_t))-\gamma \dot{y}_t + \sqrt{2 k_BT\gamma}\eta(t),\\
\langle \eta(t)\rangle &=0, \qquad \langle \eta(t),\eta(0)\rangle = \delta(t).
\end{align}
\end{subequations}
For $U(x)=k x^2/2$ and $\alpha(x) =  x$ we recover eq.~\eqref{eq_stochastic_1}.
Using eq.~\eqref{eq_stochastic_2}, 
we perform  simulations for the parameter set  $k_BT=2.5$kJ/mol, $m=50\,$u, $m_y=2\,$u, $K=30\,$kJ/mol/nm, $\gamma=10\,$u/ps  
to generate 100 trajectories $x_t$ of 100\,ns length each. 
For the potential we choose a non-linear double-well potential $U(x)=U_0(x^2-1)^2$ with $U_0=3k_BT$, 
as shown in fig.~\ref{fig_3}A, and for the non-linear coupling potential we choose a quadratic function $\alpha(x) = \alpha_0 x^2/2$ 
with $\alpha_0=4\,$nm$^{-1}$. 
The resulting  trajectories are then used to compute via  eq.~\eqref{eq_projected_forward}  all parameters of the hybrid GLE in eq.~\eqref{eq_main_result1}, which are presented in fig.~\ref{fig_3}. 
In this calculation, the $x$-space is discretized using $N_A=200$ bins of equal length.

 The effective mass $M(A_t)$ for an observable $A_t$  that is a linear function of atomic positions
is constant \cite{glatzel_interplay_2021}, as follows directly from the fact that the velocity distribution function
factorizes for Hamiltonians of the form in eq. \eqref{eq_Hamiltonian}.
Indeed, in fig.~\ref{fig_3}A  the 
numerically obtained effective potential $U_\mathrm{eff}(x)$ defined in  eq. \eqref{eq_gen_U1} (solid blue line) 
is shown to agree perfectly  with the 
input potential  $U(x)$ (broken line) when
 both potentials are shifted so that they are zero at $x=0$. 
In fig.~\ref{fig_3}B we compare  the random force distribution obtained numerically 
via eq.~\eqref{eq_projected_forward} from the simulated
 trajectory (blue  line) with a Gaussian with 
 vanishing mean and a variance of $\sqrt{\langle \dot{x}^2_0\rangle \Gamma^p(0)}$, as predicted by  eq.  \eqref{eq_Gammap},
 and obtain very good agreement; for the comparison,  the value $\Gamma^p(0)$ is numerically extracted from the simulated trajectory. 
 Note that eq.  \eqref{eq_Gammap} does not imply that the distribution of the  random force is a pure Gaussian,
 but the data in   fig.~\ref{fig_3}B demonstrate that non-Gaussian contributions are either absent or very small. 

 In fig.~\ref{fig_3}C we show  the  memory kernel  $\Gamma^p(t)$
extracted from the simulation trajectory, the result looks qualitatively similar to the result in 
 fig.~\ref{fig_1}C for the harmonic model.
% In fig.~\ref{fig_2} we show the numerical result for the function $D(x,t)$ defined by eq.~\eqref{eq_Px_memory} for different times. 
 %The $x$-space is divided into 200 bins of equal length. As predicted in  appendix~\ref{sec_App_POM}, 
 %$D(x,t)$ vanishes for all times.  
 In fig.~\ref{fig_4}, we show the correlation function $D(x,t)$ defined
  in eq.~\eqref {eq_D} for a few different fixed  times. 
  Note that $D(x,0)$ vanishes at time 0, which is true for general 
  $A_0=A(\ve R_0)$, since the product $\dot{A}_0 F^R(0)$ is odd in the momenta and
   thus the conditional average $\langle \dot{A}_0, F^R(0)\rangle_{A} = D(A, 0)$ vanishes. 
 For finite time,  $D(x,t)$ in fig.~\ref{fig_4} rises before dropping back to zero in the long-time limit. 
 The time after which $D(x,t)$ decays to zero is about 1 ps and thus comparable to the memory time of $\Gamma^p(t)$ in fig.~\ref{fig_3}C. 
Note that a finite correlation function $D(x,t)$ will  via  eq. \eqref{eq_Gammax}
give rise to a non-linear friction memory function $\Gamma^x(x,t)$.
The non-linear Zwanzig model defined by the Hamiltonian eq.~\eqref{eq_Zwanzig_model_local} is thus represented by a
constant effective mass term $M(A)$ but a non-vanishing non-linear friction memory. 

\begin{figure}
\includegraphics[width=0.8\linewidth]{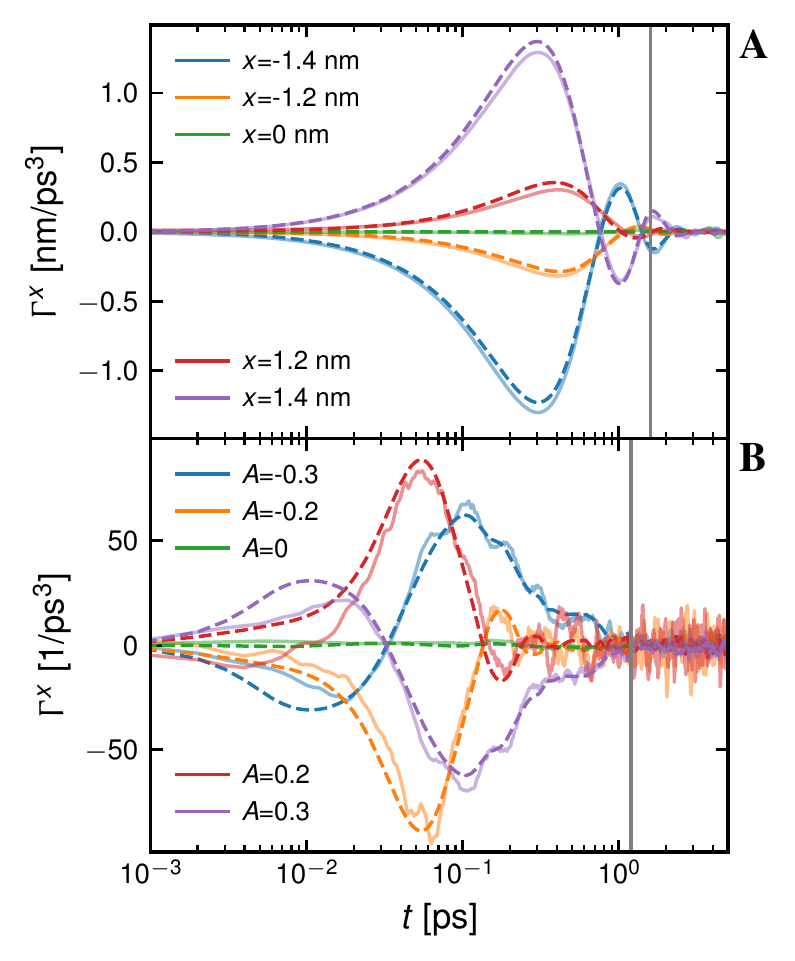}
\caption{\label{fig_6}
Non-linear friction function 
$\Gamma^x(A,t)$ as a function of time for different values of the reaction coordinate. 
\textbf{A} Results for the non-linear Zwanzig model in eq.~\eqref{eq_Zwanzig_model_local},
here  the reaction coordinate is given by the position of the relevant particle coordinate $x$.
 \textbf{B} Results  for the rescaled dihedral angle of butane from MD simulations,
 here  the reaction coordinate is given by the rescaled dihedral angle $A$.
  The dihedral angle data is more noisy compared to the non-linear Zwanzig model system because 
 of the reduced simulation time. 
 The vertical gray lines indicate the time after which the corresponding linear friction kernel 
 $\Gamma^p$ stays below 1\% of its initial value. 
 The dashed lines are obtained from a  smoothing procedure, see appendix~\ref{sec_App_smoothing}.}
\end{figure}

\subsection{Dihedral Angle Dynamics of Butane from MD Simulations
\label{sec_Apl_dihedral}}
To test our algorithm for  an observable  that is a non-linear function of atomic positions, 
we consider the dihedral angle dynamics of a butane molecule in water as obtained from MD simulations. 
The dihedral angle $\phi$ of butane is a conceptually simple yet  relevant observable and provides  a simple scenario to study
conformational transitions in polymers and proteins that is both theoretically \cite{chandler_statistical_1978} 
and experimentally \cite{zheng_ultrafast_2006} accessible. 
In fig. \ref{fig_5}, we present results for the rescaled angle $A=\phi/(\phi_\mathrm{max}-\phi_\mathrm{min})$, where 
the maximal and minimal observed angles in the studied trajectory are $\phi_\mathrm{max}=155^\circ$ and $\phi_\mathrm{min}=-157^\circ$. 
In fig. \ref{fig_5}A,
the effective potential $U_\mathrm{eff}(A)$ defined in  eq. \eqref{eq_gen_U1}  (blue solid line)  shows small 
but significant deviations from the PMF $U_\mathrm{PMF}(A)$ (broken line), which is explained by the dependence 
of the effective mass $M(A)$ on the dihedral angle, as  shown in  fig.~\ref{fig_5}D.  
In fig.~\ref{fig_5}C, the deviations between $\Gamma^p(t)$, as defined within the exact GLE eq. \eqref{eq_main_result1}
and determined numerically from the MD trajectory via eq.~\eqref{eq_projected_forward}, 
and $\Gamma^\mathrm{app}(t)$, defined within  the approximate
 GLE in eq.~\eqref{eq_GLE_adhoc} and obtained via a Volterra scheme \cite{daldrop_butane_2018, ayaz_non-markovian_2021}, 
 are pronounced and already suggest that non-linear friction effects, not captured by $\Gamma^\mathrm{app}(t)$, are present.
A closer look at the results in fig.~\ref{fig_5}C reveals that  $\Gamma^p(t)$ and $\Gamma^\mathrm{app}(t)$ have similar decay times, 
but $\Gamma^\mathrm{app}(t)$ oscillates in time while $\Gamma^p(t)$ does not. 
These deviations between $\Gamma^p(t)$ and $\Gamma^\mathrm{app}(t)$ must be due to non-linear memory effects,
 as confirmed in 
fig.~\ref{fig_5}E, where the correlation function  $D(A,t)$  defined
  in eq.~\eqref {eq_D} is shown for a few different fixed times.
 Thus, a non-linear memory friction contribution $\Gamma^x(A,t)$, defined in eq.~\eqref{eq_Gammax} and shown in fig.~\ref{fig_5}B,
 is present in the GLE. 
  As mentioned before, from the definition of $D(A,t)$ in eq.~\eqref{eq_D} it follows that $D(A,t)$ vanishes at time $t=0$, 
  i.e., $D(A,0)=0$, from which it is easy to see via eq. \eqref{eq_Gammax} that $\Gamma^x(A,0)$ vanishes, too,
  as indeed confirmed by the data in  fig.~\ref{fig_5}B and E.
For finite time, both $D(A,t)$ and $\Gamma^x(A,t)$ rise in amplitude before decaying to zero after a time corresponding
 to the memory time of $\Gamma^p(t)$ in fig.~\ref{fig_5}C, which is about 1 ps. 
 
 The rise and decay of non-linear friction effects is presented in fig.~\ref{fig_6},
  where we show $\Gamma^x(A,t)$ as a function of time for different fixed values of $A$
   for the non-linear Zwanzig model in fig.~\ref{fig_6}A and for the butane dihedral angle dynamics in fig.~\ref{fig_6}B. The vertical gray lines indicate the time after which the linear friction kernel $\Gamma^p(t)$ 
 for each system stays below 1\% of its initial value $\Gamma^p(0)$.
In addition to the raw numerical data (solid lines), we show smoothed curves
which are obtained by fits to Legendre polynomials (broken lines), 
as described in appendix~\ref{sec_App_smoothing}.

\section{Summary and Discussion}

By using a  hybrid projection scheme that combines
 linear Mori projection on the reaction coordinate velocities  and non-linear conditional Zwanzig projection on the reaction coordinates themselves, we derive a GLE that contains the non-linear  potential of mean force and a non-linear  memory friction contribution  that is a function of the 
reaction coordinate  $A_t$ but not of its velocity $\dot A_t$. The complete memory friction then splits into two parts. One part is linear in the reaction coordinate velocity
and reflects   linear friction proportional to a memory kernel  $\Gamma^p(t)$.
The  memory kernel  $\Gamma^p(t)$  is related to the fluctuating force $F^R(t)$, defined in eq.~\eqref{eq_F_operator}, 
by a relation that resembles a fluctuation-dissipation theorem, eq.~\eqref{eq_Gammap}. 
The non-linear  memory friction  function $\Gamma^x(A_{t-s},s)$ accounts for non-linear 
dependencies of friction on $A_{t-s}$ and is connected to the fluctuating force $F^R(t)$ by a conditional  correlation function, given in eq.~\eqref{eq_D}. 
Thus, when modeling $F^R(t)$ as a stochastic variable, it has to fulfill both relations, eq.~\eqref{eq_Gammap} and eq.~\eqref{eq_D}.
The approximate GLE in eq.~\eqref{eq_GLE_adhoc} is obtained from our GLE in eq.~\eqref{eq_main_result1} only when the memory friction function
$\Gamma^x(A_{t-s},s)$  vanishes, which thus establishes a firm criterion for the validity of  the approximate  GLE. 
 %We obtain our results without any assumptions on the form of the memory functions or the dependencies of the fluctuating force $F^R(t)$. The main problem with the projection operator method is that it introduces so called orthogonal dynamics into the equations of motion. The dynamics of $F^R(t)$ is generated by the projected propagator $e^{tQL}$, as given in eq.~\eqref{eq_F_operator}. Often, computing $F^R(t)$ in eq.~\eqref{eq_main_result1} in the presence of a non-linear potential force required the ad-hoc assumption that the correction term due to $\Gamma^x(A,t)$ vanishes. 
 
 We also 
  introduce a numerical scheme to compute all parameters of  our GLE from a given trajectory $ A_t$ and  apply it  on numerically determined 
 trajectories for a harmonic and a non-linear exactly solvable many-body particle system, here we  show that the numerical results agree with the analytical predictions. 
 We also apply our numerical scheme on a dihedral angle trajectory of butane in water, obtained from atomistic MD simulations. 
 We find that  the effective mass of the dihedral angle depends on the value of the dihedral angle
 and that the non-linear memory friction contribution is finite and non-negligible. 
In order to estimate the importance of the non-linear memory friction, we have to compare the linear memory kernel $\Gamma^p(t)$ and the non-linear memory function $\Gamma^x(A,t)$. For this we  multiply the linear-friction memory kernel at time zero,
 $\Gamma^p(0)$, by the root mean square velocity and obtain
  $\Gamma^p(0)\sqrt{\langle \dot{A}^2\rangle} = 171\,$ps$^{-3}$, which can be directly compared with the maximal value of the 
  non-linear memory friction function   $ \Gamma^x(A_\mathrm{max},t_\mathrm{max})=97\,$ps$^{-3}$,
  which is obtained for  $A_{max}=0.26$ and $t_\mathrm{max}=0.043$ ps. The value of $ \Gamma^x(A_\mathrm{max},t_\mathrm{max})$ thus turns out to be
   roughly half the 
  value of  $\Gamma^p(0) \sqrt{\langle \dot{A}^2\rangle}$, which means that non-linear memory friction effects are not negligible. 
 Interestingly, our results demonstrate  that non-linear friction memory  leads to oscillations in the memory function
 $\Gamma^\mathrm{app}(t)$ of the approximate  GLE, which are not present in $\Gamma^p(t)$, as shown in  fig.~\ref{fig_5}C. 
 Finally, we show that the random force in the GLE from our hybrid projection scheme exhibits  small but detectable deviations from a Gaussian distribution.
 All these results lead us to conclude that the GLE derived from our  hybrid projection scheme is practically useful and allows to detect and model 
 non-linear friction effects that have been neglected in previous applications of the approximate  GLE with linear memory friction. 
 
%The memory kernel $\Gamma^x$ in eq.~\eqref{eq_Px_memory}, which accounts for spatial memory effects, has a structure that comes up in the context of non-linear Langevin equations with multiplicative noise. One considers the non-linear Langevin equation
%\begin{align}
%\dot{x}_t &= h(x_t)+g(x_t)\eta(t),
%\end{align}
%where $\eta(t)$ is white noise. The question, which functions $h(x_t), g(x_t)$ give the correct equilibrium distribution (Boltzmann distribution), leads to the answer \cite{avalos_phenomenological_1995, risken_fokker_1996}
%\begin{align}
%\label{eq_nonlinear_Langevin}
%h(x_t) &= \frac{\mathrm{d}D}{\mathrm{d}x_t}-\beta D(x_t)\frac{\mathrm{d}U_\mathrm{PMF}}{\mathrm{d}x_t},
%\end{align}
%where $D$ is now the diffusion coefficient. An open problem remains the mapping of the hybrid GLE in eq.~\eqref{eq_main_result1} on a system of coupled Langevin equations so that one can perform efficient computer simulations of the coarse-grained model. The answer to that may be hidden in the similarity between $\Gamma^x$ in eq.~\eqref{eq_Px_memory} and the function $h$ in eq.~\eqref{eq_nonlinear_Langevin}.

\section{Methods and Materials}
MD simulations are performed using the Gromacs MD package (version 2020-Modified) \cite{pronk2013gromacs}.  For the MD simulation of the butane molecule,  we use the GROMOS53A6 force field \cite{Oostenbrink_2004}  with  the TIP4P/2005 rigid water model \cite{Abascal_2005}. The simulation box has side lengths of 3.35 nm and contains 1250 water molecules. We constrain the butane bond lengths and angles using the SHAKE algorithm \cite{Ryckaert_1977}. For long-range electrostatic interactions, we use the particle-mesh Ewald \cite{Darden_1993}, with a cut-off of 1 nm. The simulation time step is 1 fs,  and the total simulation time is 100 ns.  All simulations are performed in the NVT ensemble with a temperature of 300 K, controlled with a velocity rescaling thermostat \cite{bussi2007canonical}. Input files of the MD simulations are available upon request.
% under (\textcolor{red}{FU Refubium Link}). 
The Langevin simulations are performed using the Leap Frog algorithm for numerical integration. 
Our Python codes for extracting the GLE parameters and running Langevin simulations are also available upon request.
% in GitHub (\textcolor{red}{Link}). 
When computing non-linear memory contributions, 
the time resolution of the trajectory and the number of bins in reaction-coordinate 
space have to be chosen with care. In  our analysis of Langevin and MD simulations, we use 200 bins to discretize the reaction-coordinate  space.
For the butane dihedral angle system in fig.~\ref{fig_5}, we exclude boundary regions in the trajectory for $|A|>0.4$ in the computation of conditional correlations, that means we exclude
 observable values with a small fraction of realizations along the trajectory, since these would lead to significant noise in the
 extracted memory functions and thus destabilize the numerical extraction. Such noise effects
 are  clearly visible in the effective mass profile in fig.~\ref{fig_5}F.

\begin{acknowledgments}
We acknowledge support by Deutsche Forschungsgemeinschaft Grant CRC 1114 "Scaling Cascades in Complex System", 
Project 235221301, Project B03 
and by the ERC Advanced Grant 835117 NoMaMemo.
 We gratefully acknowledge  computing time on the HPC clusters at the physics department and ZEDAT, FU Berlin.
\end{acknowledgments}

\appendix

\section{Derivation of Eq.~\eqref{eq_cond_avrg_rel}\label{sec_App_eq13}}
In the following, we derive eq.~\eqref{eq_cond_avrg_rel} by using the definition of conditional  correlation functions
 in eq.~\eqref{eq_conditional_avrg},  the  relations in eq.~\eqref{eq_Liouville_anti_sa}, eq.~\eqref{eq_Liouville_on_delta} 
 and the definition of the PMF in eq.~\eqref{eq_def_PMF}.  We start with
\begin{subequations}
\begin{align}
\langle LB_{t^\prime}\rangle_{A_t} = \frac{\langle\delta(A(\widehat{\omega}_0)-A(\omega_t)),LB(\widehat{\omega}_0,t^\prime)\rangle}{\langle\delta(A(\widehat{\omega}_0)-A(\omega_t))\rangle},\\
=\int_{-\infty}^{\infty}\mathrm{d}a\,\delta(A(\omega_t)-a)\frac{\langle\delta(A_0-a),LB_{t^\prime}\rangle}{\langle\delta(A_0-a)\rangle},\label{eq_AppA_der}
\end{align}
where the average is over variables with a hat and consequently, the Liouville operator $L$ only acts on variables with a hat. We consider the rightmost term in eq.~\eqref{eq_AppA_der}
\begin{align}
\frac{\langle\delta(A_0-a),LB_{t^\prime}\rangle}{\langle\delta(A_0-a)\rangle} &=
 -\frac{\langle L\delta(A_0-a),B_{t^\prime}\rangle}{\langle\delta(A_0-a)\rangle}\\
&=\frac{\langle \dot{A}_0\frac{\mathrm{d}}{\mathrm{d}a}\left[\delta(A_0-a)\right],B_{t^\prime}\rangle}{\langle\delta(A_0-a)\rangle},\label{eq_AppA_der1}
\end{align}
where we used eq.~\eqref{eq_Liouville_anti_sa} and eq.~\eqref{eq_Liouville_on_delta}. We next
 pull out the derivative w.r.t. $a$ in eq.~\eqref{eq_AppA_der1} from the inner product and  use the product rule of differentiation,
 which  gives
\begin{align}
&\frac{\langle\delta(A_0-a),LB_{t^\prime}\rangle}{\langle\delta(A_0-a)\rangle} =
\frac{\mathrm{d}}{\mathrm{d}a}\frac{\langle \delta(A_0-a)\dot{A}_0,B_{t^\prime}\rangle}{\langle \delta(A_0-a)\rangle}\nonumber\\
& \qquad  + \frac{\langle \delta(A_0-a)\dot{A}_0,B_{t^\prime}\rangle}{\langle \delta(A_0-a)\rangle}\frac{\mathrm{d}}{\mathrm{d}a}\ln\langle \delta(A_0-a)\rangle.\label{eq_AppA_der2}
\end{align}
Finally, we use the definition of the PMF  in eq.~\eqref{eq_def_PMF}  and insert eq.~\eqref{eq_AppA_der2} into eq.~\eqref{eq_AppA_der} to obtain eq.~\eqref{eq_cond_avrg_rel}.
\end{subequations}

\section{Multi-Dimensional Hybrid GLE \label{sec_App_multidim}}
Here, we derive eq.~\eqref{eq_main_result1} for a multidimensional observable that is a function of particle positions only. 
We denote the set of observables using the vector $\mathbf{A}(\ve R_t) = (A_1(\ve R_t), A_2(\ve R_t), \dots, A_n(\ve R_t))$. As before, all observables implicitly depend on time via the positions $\ve R_t$. 
We denote components as  $A_k(\ve R_t) \equiv A_{k,t}$ and $A_k(\ve R_0) \equiv A_{k,0}$. In the multi-dimensional case, the projection operator reads
for general vectorial projection function $\mathbf{B}_0$ 
\begin{align}
P_H A_{m,t} & = (P_p + P_x)A_{m,t},\nonumber\\
&=  \sum_{k=1}^{n} \frac{\langle A_{m,t}, \dot{B}_{k,0}\rangle}{\langle \dot{B}_{k,0}^2\rangle}\dot{B}_{k,0}
+\langle A_{m,t} \rangle_{\mathbf{B}_0}.\label{eq_gen_projection}
\end{align}
Choosing $\mathbf{B}_0  = \mathbf{A}_0$, as in the main text, 
the projection in eq.~\eqref{eq_gen_projection} leads to the following potential term
\begin{align}
&e^{tL} P_x L \dot{\mathbf A}_0 = k_B T\, \left(\nabla^T_{A}\cdot M^{-T}(\mathbf A_t)\right)^T\nonumber\\
 & \qquad - M^{-1}(\mathbf{A}_t)\cdot\nabla_A U_\mathrm{PMF}(\mathbf{A}_t).\label{eq_effective_force}
\end{align}
where we introduced the  inverse generalized mass matrix
\begin{align}
M^{-1}_{kl}(\mathbf A) &= \beta \langle \dot{A}_{k,0},\dot{A}_{l,0}\rangle_{\mathbf{A}}. \label{eq_gen_mass}
\end{align}
The computation of the memory function proceeds similarly as in the main text and the multi-dimensional  GLE reads
\begin{align}
\label{eq_gen_GLE}
\ddot{\mathbf A}_t &=  k_B T\, \left(\nabla^T_{A}\cdot M^{-T}(\mathbf A_t)\right)^T
  - M^{-1}(\mathbf{A}_t)\cdot\nabla_A U_\mathrm{PMF}(\mathbf{A}_t) \nonumber\\
& - \int_0^t\mathrm{d}s\,\Gamma^p(t-s)\cdot \dot{\mathbf A}_s + \int_0^t\mathrm{d}s\,\boldsymbol\Gamma^x(\mathbf{A}_{t-s},s)
 + \mathbf{F}^R(t),
\end{align}
where the following relations hold
\begin{align}
&\langle F^R_k(t)\rangle = 0,  \qquad \langle F^R_k(t), F^R_l(0)\rangle = \langle \dot{A}_0^2\rangle\, \Gamma^p_{kl}(t),\nonumber\\
&\langle F^R_k(t), A_{l,0}\rangle =0, \qquad \langle F^R_k(t),\dot{A}_{l,0}\rangle =0,
\end{align}
for all $k,l=1,2,\dots,n$. The $k$-th component of the vectorial  non linear memory 
memory friction function $\boldsymbol\Gamma^x(\mathbf{A},s)$ is given by
\begin{subequations}
\begin{align}
&\Gamma^x_k (\mathbf{A}, s) = P_x L F^R_k(s)\nonumber\\ 
& = \left[ \nabla_A\cdot\mathbf{D}_k(\mathbf{A},s)-\beta\, \mathbf{D}_k(\mathbf{A},s)\cdot \nabla_A U_\mathrm{PMF}(\mathbf{A})\right],\label{eq_gen_Gammax}\\
&\mathbf{D}_k(\mathbf{A}, s) = \langle \dot{\mathbf{A}}_0,\, F^R_k(s)\rangle_{\mathbf{A}}.\label{eq_gen_D}
\end{align}
\end{subequations}

\section{Idempotency of the Hybrid Projection Operator \label{sec_App_idem}}
The linear operator $P_H$  in eq.~\eqref{eq_projection} is a projection, if it is idempotent, i.e., $P_H^2=(P_p+P_x)^2=P_H$. Clearly, we have $P_p^2=P_p$ and $P_x^2=P_x$. Therefore, one has to check that $P_pP_x\, A_t = P_xP_p\, A_t= 0$ for an arbitrary observable $A_t=A(\omega_t)$. This is true because of the following: we project onto observables of positions only, i.e., onto $B_0=B(\ve R_0)$. 
Thus, the velocity $\dot{B}_0$ is linearly proportional to the  particle  momenta
\begin{align}
\dot{B}_0 = L B_0 = \sum_{n=1}^N \frac{\mathbf{p}_n}{m_n}\cdot\nabla_{r_n}B_0.\label{eq_AppB_der1}
\end{align}

The operator $P_p$ maps any function onto the subspace of functions that are linear in the observable velocity $\dot{B}_0$, 
which is linear in the particle  momenta $\mathbf{p}_n$. 
From this we see that
\begin{align}
P_x P_p\, A_t &\propto P_x \dot{B}_{0} = 0,
\end{align}
since the operator $P_x$ involves an integral over the particle momenta but adds no momentum dependence.

$P_x$ maps any observable onto a function which depends on particle positions only. Since 
 $\dot{B}_0$ is linearly proportional to the particle momenta, $P_p$ applied on a function that depends on particle positions only 
 gives zero. Therefore, it follows  that
\begin{align}
P_p P_x A_t & = 0.
\end{align} 

\section{Self-Adjointedness and Orthogonality of Hybrid Projection \label{sec_App_ortho}}
Here, we prove that the projection $P_H$ in eq.~\eqref{eq_projection} is self-addjoint  w.r.t. the inner product in eq.~\eqref{eq_inner_product}, i.e., for any observables $A_t=A(\omega_t)$ and $C_{t^\prime}=C(\omega_{t^\prime})$, 
we have $\langle A_t, P_H C_{t^\prime}\rangle = \langle P_H A_t, C_{t^\prime}\rangle$. 
For this, we consider the projection operators $P_p$ and $P_x$ separately.

Using the definition in eq.~\eqref{eq_projection_p}, we find
\begin{subequations}
\begin{align}
\left\langle A_t, P_p C_{t^\prime}\right\rangle &= \left\langle A_t,\frac{\langle \dot B_{0}, C_{t^\prime}\rangle}{\langle \dot B_{0}^2\rangle} \dot B_{0}\right\rangle\\
 &= \langle A_t, \dot B_{0}\rangle\frac{\langle \dot B_{0}, C_{t^\prime}\rangle}{\langle \dot B_{0}^2\rangle}\\
&= \left\langle \frac{\langle A_t, \dot B_{0}\rangle}{\langle \dot B_{0}^2\rangle} \dot B_{0}, C_{t^\prime}\right\rangle\\
&= \langle P_p A_t, C_{t^\prime}\rangle.
\end{align}
\end{subequations}

Using the definition in eq.~\eqref{eq_projection_x} and eq.~\eqref{eq_conditional_avrg}, we find
\begin{subequations}
\begin{align}
&\langle A_t, P_x C_{t^\prime}\rangle\\
&= \left\langle A(\omega^\prime_t), \int\mathrm{d}a\,\delta(B(\ve R^\prime_0)-a) \frac{\langle \delta(B(\widehat{\ve R}_0) -a), C(\widehat{\omega}_{t^\prime})\rangle}{\mathbb{P}(a)}\right\rangle\nonumber\\
&= \int\mathrm{d}a\,\langle A(\omega^\prime_t),\delta(B(\ve R^\prime_0) - a)\rangle \frac{\langle \delta(B(\widehat{\mathbf{R}}_0) -a), C(\widehat{\omega}_{t^\prime})\rangle}{\mathbb{P}(a)}\nonumber\\
&= \left\langle \int\mathrm{d}a\, \frac{\langle A(\omega^\prime_t),\delta(B(\mathbf{R}^\prime_0) - a)\rangle}{\mathbb{P}(a)} \delta(B(\widehat{\mathbf{R}}_0) -a), C(\widehat{\omega}_{t^\prime})\right\rangle\nonumber\\
&= \langle P_x A_t, C_{t^\prime}\rangle.
\end{align}
\end{subequations}
This means that the hybrid projection $P_H$ in eq.~\eqref{eq_projection} is self-adjoint and thus is an orthogonal projection, i.e.,
\begin{align}
\langle P_H A_t, Q_H C_{t^\prime}\rangle =0,
\end{align}
for arbitrary observables $A_t$ and $C_{t^\prime}$.

\section{Average of Complementary Observables Vanishes \label{sec_App_equil_avrg}}
In the following, we prove eq.~\eqref{eq_avereage_orthogonal_dyn}, i.e., we show that the equilibrium average of any observable that lies in the complementary subspace at all times vanishes. For this, we must show for an arbitrary observable 
$A(\omega_t)=A_t$ that $\langle P_H A_t\rangle = \langle A_t\rangle$ holds. 
First, from the definition of $P_p$ in eq.~\eqref{eq_projection}, it follows that
\begin{align}
\langle P_p A_t\rangle &\propto \langle \dot B_{0}\rangle = 0,
\end{align}
since our projection function $B_0=B(\mathbf{R}_0)$ is a function of positions only and therefore, 
its velocity $\dot{B}_0=LB_0$ is linear in the momenta (see eq.~\eqref{eq_AppB_der1}). For 
the $P_x$ projection operator we find 
\begin{subequations}
\begin{align}
 \langle P_x A_t\rangle &= \langle \langle A_t\rangle_{B_0}\rangle = \langle A(t)\rangle.
\end{align}
\end{subequations}
From this, it immediately follows that
$\langle P_H A_t\rangle = \langle A(t)\rangle$ and thus 
 all equilibrium averages in the complementary  subspace vanish, i.e., 
 $\langle Q_H A_t\rangle = \langle (1-P_H) A_t\rangle =0$. In particular, the equilibrium average of the random force vanishes at all times, 
 i.e., $\langle F(t)\rangle = \langle Q_H F(t)\rangle = 0$.

\section{Generalized Mass of Distance Observables  \label{sec_App_Mass}}
 We demonstrate that the generalized mass $M(A)$ defined in eq.~\eqref{eq_gen_M} 
 is constant for an observable that corresponds to the scalar distance between particle positions,
 which is a non-linear function of particle positions. 
 In this case,  the force term $\mathrm dU_\mathrm{eff} /\mathrm d A$ in eq.~\eqref{eq_main_result1} reduces to $\mathrm{d}U_\mathrm{PMF}/\mathrm{d}A$. 
 As an example, we  consider 
 the hydrogen-bond distance between a nitrogen atom (donor) with initial position $\mathbf{r}^N_0$ and 
 an oxygen atom (acceptor) with initial position $\mathbf{r}^O_0$ that are located four residues apart  on
  the backbone of a polypeptide. The observable is thus given by 
\begin{align}
\label{eq_distance}
A_0 = A(\mathbf{R}_0) &= \sqrt{(\mathbf{r}^N_0-\mathbf{r}^O_0)^2} .
\end{align}
Applying the Liouville operator on eq.~\eqref{eq_distance} gives the velocity of the observable
\begin{align}
\label{eq_dist_vel}
LA_0 &= \dot{A}_0 = \left(\frac{\mathbf{p}^N_0}{m_N}-\frac{\mathbf{p}^O_0}{m_O}\right)\cdot\frac{(\mathbf{r}^N_0-\mathbf{r}^O_0)}{A_0}.
\end{align}
As can be seen in eq.~\eqref{eq_dist_vel}, the velocity $\dot{A}_0$ is linear in the momenta $\mathbf{p}^N_0$ and $\mathbf{p}^O_0$. 
Computing the effective mass according to the definition in eq.~\eqref{eq_gen_M}, i.e.,
\begin{align}
\label{eq_AppE_mass}
\langle \dot{A}_0^2\rangle_{A_0} &= \frac{k_BT}{M(A_0)} = \frac{\langle \delta [A(\widehat{\mathbf{R}}_0)-A_0], \dot{A}(\widehat{\mathbf{R}}_0)^2\rangle}{\langle \delta [A(\widehat{\mathbf{R}}_0)-A_0]\rangle},
\end{align}
requires the computation of the numerator on the r.h.s. of eq.~\eqref{eq_AppE_mass}. Given an Hamiltonian of the form in eq.~\eqref{eq_Hamiltonian}, 
factorization of the phase-space integral leads to
\begin{align}
\label{eq_AppE_mass2}
&\langle \delta [A(\widehat{\mathbf{R}}_0)-A_0], \dot{A}(\widehat{\mathbf{R}}_0)^2\rangle\nonumber\\
& \qquad = k_BT\left(\frac{1}{m_N}+\frac{1}{m_O}\right) \langle \delta [A(\widehat{\mathbf{R}}_0)-A_0]\rangle.
\end{align}
Inserting eq.~\eqref{eq_AppE_mass2} for the numerator on the r.h.s. of eq.~\eqref{eq_AppE_mass}, we find 
\begin{align}
\label{eq_AppE_mass3}
M(A_0) = M = \frac{m_N\, m_O}{m_N+m_O},
\end{align}
which is the reduced mass of the nitrogen-oxygen distance coordinate. 

A similar derivation can also be done for a linear combination of distances. For example, consider the mean hydrogen-bond distance between 
$N_R$ donor nitrogen atoms and $N_R$ acceptor oxygen atoms that are located four residues apart along the backbone of a polypeptide. 
We define  the observable as
\begin{align}
A_0 = \frac{1}{N_R}\sum_{n=1}^{N_R} A_{n,0},
\end{align}
with $A_{n,0}$ being the initial value of the $n$-th distance. Eq.~\eqref{eq_AppE_mass} becomes
\begin{align}
\langle \dot{A}_0^2\rangle_{A_0} =
 \left \langle \left( \frac{1}{N_R} \sum_{n=1}^{N_R} \dot A_{n,0}\right)^2 \right \rangle_{A_0} =
  \frac{1}{N_R^2}\sum_{n=1}^{N_R}\langle \dot A_{n,0}^2\rangle_{A_0}.
\end{align}
As before, terms consisting of mixed momentum factors  average to zero, only diagonal terms contribute. 
In analogy to eq.~\eqref{eq_AppE_mass3}, the effective mass is constant also for this case. 

\section{Alternative Discretization of Eq.~\eqref{eq_projected_4}\label{sec_App_trapz}}
Here, we present an alternative discretization of eq.~\eqref{eq_projected_4}. 
Similar to eq.~\eqref{eq_projected_forward}, the equation derived here  still has
 an overall error of the order $\mathcal{O}(\Delta t^2)$.  
 The advantage over eq.~\eqref{eq_projected_forward} is that we use the trapezoidal rule for the integration involving the memory kernel $\Gamma^p(t)$; note that we  keep the rectangular rule for the integration of the memory function $\Gamma^x(A,t)$. 
 We discretize eq.~\eqref{eq_projected_4} in the following way
\begin{align}
\label{eq_trapz}
F^R(i,j+1) &= F^R(i+1, j)\nonumber\\
& + \frac{\Delta t}{2}\, \Gamma^p(j) \dot{A}_{i+1} + \frac{\Delta t}{2}\, \Gamma^p(j+1) \dot{A}_{i}\nonumber\\
& - \Delta t\, \Gamma^x(A_{i+1},j) + \mathcal{O}(\Delta t^2).
\end{align}
Note that now, the r.h.s. of eq.~\eqref{eq_trapz} depends on $\Gamma^p(j+1)$. To compute $\Gamma^p(j+1)$, we need the yet unknown $F^R(i,j+1)$. In the absence of the memory function $\Gamma^x(A,t)$, it has been demonstrated how one can work around this problem \cite{lesnicki_molecular_2016,jung_iterative_2017,klippenstein_cross-correlation_2021}. The trick is to multiply eq.~\eqref{eq_trapz} by $F^R(i,0)$ and average according to eq.~\eqref{eq_inner_product}. This gives
\begin{align}
\label{eq_trapz2}
&\langle F^R(i,j+1), F^R(i,0)\rangle = \langle F^R(i+1, j), F^R(i,0)\rangle\nonumber\\
& + \frac{\Delta t}{2}\, \Gamma^p(j) \langle \dot{A}_{i+1}, F^R(i,0)\rangle + \frac{\Delta t}{2}\, \Gamma^p(j+1) \langle \dot{A}_{i}, F^R(i,0)\rangle\nonumber\\
& - \Delta t\, \langle \Gamma^x(A_{i+1},j), F^R(i,0)\rangle.
\end{align}
By identifying the l.h.s. of eq.~\eqref{eq_trapz2} with $\langle \dot{A}_0^2\rangle \Gamma^p(j+1)$ and solving 
for $\Gamma^p(j+1)$, we find
%\begin{align}
%&\left( 1 - \frac{\Delta t}{2} \frac{\langle \dot{A}_{i}, F^R(i,0)\rangle}{\langle\dot{A}_0^2\rangle}\right)\Gamma^p(j+1) = \frac{\langle F^R(i+1, j), F^R(i,0)\rangle}{\langle\dot{A}_0^2\rangle}\nonumber\\
%& + \frac{\Delta t}{2}\, \Gamma^p(j) \frac{\langle \dot{A}_{i+1}, F^R(i,0)\rangle}{\langle\dot{A}_0^2\rangle}\nonumber\\
%& - \Delta t\, \frac{\langle \Gamma^x(A_{i+1},j), F^R(i,0)\rangle}{\langle\dot{A}_0^2\rangle}.
%\end{align}
\begin{subequations}
\label{eq_trapz2b}
\begin{align}
\Gamma^p(j+1) &= \frac{\xi(j)+\frac{\Delta t}{2}\Gamma^p(j)\,\zeta-\Delta t\, \eta(j)}{1-\frac{\Delta t}{2}\chi},\\
\xi(j) &= \frac{\langle F^R(i+1, j), F^R(i,0)\rangle}{\langle\dot{A}_0^2\rangle},\\
\zeta &= \frac{\langle \dot{A}_{i+1}, F^R(i,0)\rangle}{\langle\dot{A}_0^2\rangle},\\
\eta(j) &= \frac{\langle \Gamma^x(A_{i+1},j), F^R(i,0)\rangle}{\langle\dot{A}_0^2\rangle},\label{eq_AppG_eta}\\
\chi &= \frac{\langle \dot{A}_{i}, F^R(i,0)\rangle}{\langle\dot{A}_0^2\rangle}.
\end{align}
\end{subequations}
The function $\eta(j)$ in eq.~\eqref{eq_AppG_eta} appears due to the presence of the non-linear friction $\Gamma^x(A,t)$ and is computed using
\begin{subequations}
\label{eq_trapz3}
\begin{align}
&\eta(j) = \int_{-\infty}^{\infty}\mathrm{d}a\,\frac{\langle\delta\left(A_{i+1}-a\right), F^R(i,0)\rangle}{\langle \dot{A}_0^2\rangle}\, \Gamma^x(a,j),\\
&\langle\delta\left(A_{i+1}-a\right), F^R(i,0)\rangle = \frac{\displaystyle \sum_{\substack{0\leq i\leq N_{traj}-1\\ A_{i+1}\in I_a}} F^R(i,0)}{\displaystyle \sum_{\substack{0\leq i\leq N_{traj}-1\\ A_{i+1}\in I_a}} 1}.
\end{align}
\end{subequations}
The alternative discretization is then found by replacing $\Gamma^p(j+1)$ on the r.h.s. of eq.~\eqref{eq_trapz} by eq.~\eqref{eq_trapz2}.

\section{Solving the Harmonic Hamiltonian Model Using Hybrid Projection \label{sec_App_POM}}
We  derive the GLE for the harmonic Hamiltonian model eq.~\eqref{eq_Zwanzig_model_nonlocal} using our 
hybrid projection in eq.~\eqref{eq_projection}. The Liouville operator defined in  eq.~\eqref{eq_Liouville} reads 
\begin{subequations}
\label{eq_model_Liouville}
\begin{align}
L &= L_x + L_q,\\
L_x &= \frac{p}{m}\frac{\partial }{\partial x} -\left( k x +\sum_{n=1}^N k_n(x-q_n) \right)\frac{\partial }{\partial p},\\
L_q &= \sum_{n=1}^N\left( \frac{v_n}{m_n}\frac{\partial }{\partial q_n} -k_n(q_n-x)\frac{\partial }{\partial v_n}\right)
\end{align}
\end{subequations}
and  acts on the  initial values $x_0, p_0, q_{n,0}, v_{n,0}$. The hybrid  projection is given by
\begin{subequations}
\label{eq_model_projection}
\begin{align}
P_H &= P_x + P_p,\\
P_p A_t &= \frac{\langle p_0, A_t\rangle}{\langle p_0^2\rangle}p_0, & P_x A_t &= \langle A_t \rangle_{x_0},\label{eq_model_Px}
\end{align}
\end{subequations}
with the conditional average in eq.~\eqref{eq_model_Px} being defined in eq.~\eqref{eq_conditional_avrg}. Using eq.~\eqref{eq_model_Liouville} and eq.~\eqref{eq_model_projection}, it follows that
\begin{align}
-U^\prime_\mathrm{PMF}(x_t) &= e^{tL}P_HLp_0 = -k x_t,
\end{align}
as confirmed  in fig.~\ref{fig_1}A. To compute the random force $F^R(t)$, we use the operator expansion
\begin{align}
F^R(t) &= e^{tQ_HL}Q_HLp_0 = \sum_{j=0}^\infty \frac{t^j}{j!}(Q_HL)^j Q_HLp_0,
\end{align}
and repeatedly apply the operator $Q_HL$ on $Q_HLp_0=\sum_{n=1}^N k_n(q_{n,0}-x_0)$. We find
\begin{align}
&F^R(t)  = \sum_{n=1}^N\left[\left( \mu_n t - \frac{(\mu_n t)^3}{3!}  + \frac{(\mu_n t)^5}{5!} + \dots \right) \mu_n v_{n,0}\right.\nonumber\\
& + \left. k_n \left( 1 - \frac{(\mu_n t)^2}{2!} + \frac{(\mu_n t)^4}{4!} + \dots \right) (q_{n,0}-x_0)\right],
\end{align}
with $\mu_n = \sqrt{k_n/m_n}$. Identifying the sums in the parenthesis as the series expansions of sine and cosine, 
respectively, $F^R(t)$ follows  as
\begin{align}
\label{eq_model_FR}
F^R(t) &= \sum_{n=1}^N\left( \mu_n \sin(\mu_n t) v_{n,0} + k_n \cos(\mu_n t) (q_{n,0}-x_0)\right).
\end{align}
The result  in eq.~\eqref{eq_model_FR} equals the result given in eq.~\eqref{eq_FR2} for the same model, obtained by 
setting  $\alpha(x)=x$,
 which follows  by explicit solution of the equations of motion. 
 Using $F^R(t)$ in eq.~\eqref{eq_model_FR} to compute $D(x,t)$, one obtains 
\begin{align}
D(x,t) &= \left\langle \frac{p_0}{m}, F^R(t)\right\rangle_{x} = 0.
\end{align}
Hence, the memory function $\Gamma^x(x,t)$ in eq.~\eqref{eq_Px_memory} vanishes. The memory function $\Gamma^p(t)$ follows as 
\begin{align}
\label{eq_AppH_Gammap}
\Gamma^p(t) =\frac{\langle F^R(t), F^R(0)\rangle}{\langle p_0^2\rangle} = \frac{1}{m}\sum_{n=1}^N k_n \cos(\mu_n t). 
\end{align}
The friction integral in the GLE reads
\begin{align}
\int_0^t\mathrm{d}s\,\Gamma^p(s) p_{t-s} = \int_0^t\mathrm{d}s\,\Gamma(s) \dot{x}_{t-s},
\end{align} 
with $\Gamma(s)=\sum_{n=1}^N k_n \cos(\mu_n s)$ being the result in eq.~\eqref{eq_mem_2} obtained by explicit solution of the equations of motion for the special case  $\alpha(x)=x$. 

\section{Derivation of a GLE for the Non-Linear Zwanzig Model\label{sec_App_ZwanzigGLE}}
We demonstrate how to derive a GLE from the Hamiltonian version of the non-linear Zwanzig model defined 
in eq.~\eqref{eq_Zwanzig_model_local} \cite{zwanzig_nonlinear_1973}. 
In the main text, we perform numerical simulations of the resulting GLE to produce trajectories on which we can test
our numerical extraction techniques. 
The coordinates of the relevant particle are given by $(x,p)$,
 a non-linear function $\alpha(x)$ enters the coupling to  the auxiliary variables $\lbrace q_n, v_n\rbrace$.
 This represents a symmetry breaking in the interactions,
 since the interactions do not depend on the distance $|q_n-x|$, 
 but rather  on the explicit value of $x$. 
 The harmonic model  defined by  eq.~\eqref{eq_Zwanzig_model_nonlocal} follows from  the non-linear model eq.~\eqref{eq_Zwanzig_model_local} 
 in the special case  $\alpha(x)=x$, so the solution of the model in  eq.~\eqref{eq_Zwanzig_model_nonlocal} is obtained by setting $\alpha(x)=x$
 in the final results obtained in this section.
 As we will show here, a non-linear $\alpha(x)$ induces non-linear memory friction in  the corresponding GLE. 
 
 The equations of motion for the Hamiltonian in eq.~\eqref{eq_Zwanzig_model_local} read
\begin{subequations}
\begin{align}
m \ddot{x}_t &= -U^\prime(x_t) -\sum_n k_n \alpha^\prime(x_t)\left(\alpha(x_t) - q_{n,t}\right),\label{eq_x1}\\
m_n \ddot{q}_{n,t} &= -k_n\left(q_{n,t} - \alpha(x_t)\right)\label{eq_x2},
\end{align}
\end{subequations}
where the prime superscript denotes a derivative w.r.t. the argument, 
i.e., $U^\prime(x) = \mathrm{d}U/\mathrm{d}x$. Eq.~\eqref{eq_x2} can be solved to give
\begin{align}
\label{eq_solx2_1}
q_{n,t} &= q_{n,0} \cos(\mu_n t) + \frac{v_{n,0}}{m_n \mu_n}\sin(\mu_n t)\nonumber\\
& \quad + \mu_n \int_0^t\mathrm{d}s\, \sin(\mu_n (t-s))\, \alpha[x_s],
\end{align}
where $\mu_n^2 = k_n/m_n$. By partial integration, the solution in eq.~\eqref{eq_solx2_1} can be written in the form
\begin{align}
\label{eq_solx2_2}
q_{n,t} &= \left(q_{n,0} - \alpha(x_0)\right) \cos(\mu_n t) + \frac{v_{n,0}}{m_n \mu_n} \sin(\mu_n t)\nonumber\\
& \quad - \int_0^t\mathrm{d}s\, \cos(\mu_n (t-s))\,  \alpha^\prime(x_s) \dot{x}_s + \alpha(x_t).
\end{align}
Inserting eq.~\eqref{eq_solx2_2} into eq.~\eqref{eq_x1} leads to a GLE for $x$, i.e.,
\begin{subequations}
\begin{align}
\label{eq_Zwanzig_GLE1}
m \ddot{x}_t &= -U^\prime(x_t) + F^R_Z(t, x_t)\nonumber\\
& \quad - \int_0^t\mathrm{d}s\,\Gamma\left[t-s,x_t,x_s\right]\dot{x}_s,\\
F^R_Z(t, x_t) &= \sum_n \alpha^\prime (x_t)\bigg(\mu_n v_{n,0} \sin(\mu_n t)\nonumber\\
& + k_n \left(q_{n,0} - \alpha(x_0)\right) \cos(\mu_n t)\bigg),\label{eq_FR2}
\end{align}
with the memory function
\begin{align}
\Gamma(t-s,x_t,x_s) & = \sum_n k_n \alpha^\prime(x_t) \alpha^\prime(x_s) \cos(\mu_n(t-s))
\label{eq_mem_2}.
\end{align}
\end{subequations}

\section{Markovian Embedding of the Non-Linear Zwanzig Hamiltonian Model \label{sec_App_StochasticZwanzig}}
Here, we show how to map the non-linear Zwanzig Hamiltonian system defined in eq.~\eqref{eq_Zwanzig_model_local} onto 
a Markovian stochastic system of equations, for which numerical simulations can be efficiently performed. 
The results obtained here include the harmonic model  in eq.~\eqref{eq_Zwanzig_model_nonlocal} by setting $\alpha(x)=x$. 
Consider the memory function in eq.~\eqref{eq_mem_2}. It contains a sum over cosines, i.e., $\sum_{n=1}^N k_n \cos(\mu_n t)$. For $N\to\infty$, this represents a Fourier series of an even, periodic function in time with Fourier coefficients $k_n$. In the  continuous limit, 
i.e., $k_n\to k(\mu)\mathrm{d}\mu/2\pi$, this defines an arbitrary even function $f(t)$  
\begin{align}
\label{eq_stochastic_mem}
\sum_{n=1}^\infty k_n &\cos(\mu_n t) \to \int_{-\infty}^\infty \frac{\mathrm{d}\mu}{2\pi}\,k(\mu) \cos(\mu t) = f(t) \\
&= K\, e^{-|t|/\tau}\left(\cos\left(\frac{2 \pi}{T} t\right)+c\sin\left(\frac{2\pi}{T} |t|\right)\right)
\end{align}
with an exponential decay time $\tau$ and parameters  $T$,  $K$, $c$ to be determined below. 
The function $k(\mu)$ follows from the Fourier transform as 
\begin{align}
k(\mu) &= \int_{-\infty}^{\infty}\mathrm{d}t\, \cos\left(\mu t\right) f(t).\label{eq_mapping}
\end{align}
 The memory function in eq.~\eqref{eq_mem_2} becomes
\begin{align}
\label{eq_mem_stochastic}
\Gamma(t-s,x_t,x_s) &  = \alpha^\prime(x_t) \alpha^\prime(x_s)\,K\, e^{-|t-s|/\tau}\nonumber\\
&\times\left(\cos\left(\frac{2\pi}{T} (t-s)\right)+c\sin\left(\frac{2\pi}{T}|t-s|\right)\right).
\end{align}
Now, consider the random force in eq.~\eqref{eq_FR2}. For $N\to\infty$, it  can  be rewritten as
\begin{align}
\label{eq_mem_Ftilde}
F^R_Z(t,x_t) &= \alpha^\prime(x_t) \tilde F^R_Z(\lbrace q_{n,0},v_{n,0}\rbrace, t).
\end{align}
In the stochastic interpretation of the GLE, it is sufficient to know the distribution of the initial conditions of
the complementary variables. 
For the Hamiltonian in eq.~\eqref{eq_Zwanzig_model_local}, the distribution is given by the Boltzmann distribution. 
Thus, the initial values   $q_{n,0}, v_{n,0}$ are Gaussian distributed random variables with
\begin{subequations}
\begin{align}
&\langle (q_{n,0}-\alpha(x_0)) \rangle = 0, \qquad \langle v_{n,0} \rangle = 0,\\
&\langle \alpha(x_0),v_{n,0}\rangle = 0, \qquad \langle q_{n,0},v_{n,0}\rangle = 0,\\
&\langle v_{n,0},v_{m,0}\rangle = \delta_{n,m}\frac{k_B T}{m_n},\\
&\langle (q_{n,0}-\alpha(x_0)),(q_{m,0}-\alpha(x_0))\rangle = \delta_{n,m} \frac{k_B T}{k_n}.\label{eq_AppJ_rel}
\end{align}
\end{subequations}
From this, it follows that $\tilde F^R_Z$ is a stationary Gaussian process fulfilling
\begin{subequations}
\begin{align}
&\langle \tilde F^R_Z(t)\rangle = 0,\\
&\langle\tilde{F}^R_Z(t),\tilde F^R_Z(0)\rangle = k_BT \sum_n k_n \cos(\mu_n t) \to k_BT f(t),\label{eq_AppJ_rel2}
\end{align}
\end{subequations}
The equal sign in eq.~\eqref{eq_AppJ_rel2} follows from the explicit form given in eq.~\eqref{eq_FR2} and from the relation in eq.~\eqref{eq_AppJ_rel}, where the average is a Boltzmann average over the initial conditions $\lbrace q_{n,0},v_{n,0}\rbrace$. 
A Markovian stochastic system which leads to a memory function
 of the form given in eq.~\eqref{eq_mem_stochastic} reads
\begin{subequations}
\label{eq_stochastic_eom}
\begin{align}
m \ddot{x}_t &= -U^\prime(x_t) - k \alpha^\prime(x_t)\left(\alpha(x_t)-y_t\right),\label{eq_stochastic_eom_a}\\
m_y \ddot{y}_t &= -k\left(y_t-\alpha(x_t)\right)-\gamma\dot{y}_t + \sqrt{2k_BT\gamma}\eta(t).\label{eq_stochastic_eom_b}
\end{align}
\end{subequations}
with $\langle \eta(t)\rangle = 0$, $\langle \eta(t),\eta(s)\rangle = \delta(t-s)$ being white noise. 
The relations between the parameters in eq.~\eqref{eq_stochastic_eom} and the parameters in eq.~\eqref{eq_stochastic_mem} are given by
\begin{subequations}
\begin{align}
&\nu^2 = 4\tau_\Gamma^2\mu^2-1,\\
&T = \frac{4\pi}{\nu}\tau_\Gamma, & &\tau = 2 \tau_\Gamma, &  &K = k,\\
&\tau_\Gamma = \frac{m_y}{\gamma}, & &\mu^2 = \frac{k}{m_y} & &c = \frac{1}{\nu}.
\end{align}
\end{subequations}
By solving eq.~\eqref{eq_stochastic_eom_b} and inserting the result into eq.~\eqref{eq_stochastic_eom_a}, we find the random force
\begin{align}
\label{FR2}
\tilde{F}^R(t) &= k e^{-|t|/2\tau}\left(\cos\left(\frac{2\pi}{T}t\right) + \frac{1}{\nu}\sin\left(\frac{2 \pi}{T} t\right)\right)\nonumber\\
& \quad \times\left(y_0-\alpha(x_0)\right)+\frac{2}{\gamma \nu} e^{-|t|/2\tau}\sin\left( \frac{2 \pi}{T} t \right)\nonumber\\
& \quad \times p_{y,0} + \sqrt{2 k_BT \gamma} \int_0^t\mathrm{d}s\,2 e^{-(t-s)/2\tau}\frac{\tau}{\nu}\nonumber\\
& \quad \times\sin\left( \frac{2 \pi}{T} (t-s) \right) \eta(s),
\end{align}
where the variable $y_0$ has the same distribution as $q_{n,0}$, and $p_{y,0}=m_y\dot{y}_0$ has the same distribution as $v_{n,0}$. The equivalence of $\tilde F^R$ in eq.~\eqref{eq_mem_Ftilde} and $\tilde F^R_Z$ in eq.~\eqref{FR2} follows from the fact that 
their first and second moments are the same. 
Using this, we have mapped the non-linear Hamiltonian Zwanzig model defined by eq.~\eqref{eq_Zwanzig_model_local} 
onto the set of coupled Markovian stochastic equations in eq.~\eqref{eq_stochastic_eom}, which can be used to  perform 
numerical simulations.

\section{Transformation Between Different  GLEs\label{sec_App_Mapping_GLEs}}

When applied to the non-linear Hamiltonian Zwanzig model  defined by eq.~\eqref{eq_Zwanzig_model_local}, 
our hybrid projection operator $P_H = P_x + P_p$, given in eq.~\eqref{eq_model_projection}, leads to a GLE of the form
\begin{subequations}
\label{eq_Zwanzig_GLE_projection}
\begin{align}
\label{eq_Zwanzig_GLE2}
\dot{p}_t &= -U^\prime(x_t) + F^R(t)\nonumber\\
& \quad -\int_0^t\mathrm{d}s\,\Gamma^p(t-s) p_s\nonumber\\
& \quad +\int_0^t\mathrm{d}s\,\Gamma^x(t-s,x_s) ,\\
\Gamma^p(t) &= \frac{\langle F^R(t),F^R(0)\rangle}{\langle p_0^2\rangle},\label{eq_app_Gammap}\\
D(s,x_{t-s})&=\left\langle \frac{p_0}{m},F^R(s)\right\rangle_{x_{t-s}},
\end{align}
\end{subequations}
where $\Gamma^x(x_{t-s},s)$ follows from eq.~\eqref{eq_Gammax}. 
The two GLE's in eq.~\eqref{eq_Zwanzig_GLE1} and eq.~\eqref{eq_Zwanzig_GLE2} obviously have a different mathematical structure, 
but they describe the exact same dynamics. To see this, consider the random force $F^R_Z(t,x_t)$ in eq.~\eqref{eq_FR2}
\begin{subequations}
\label{eq_AppK_FR}
\begin{align}
F^R_Z(t,x_t) &=  \alpha^\prime(x_t)\tilde{F}^R_Z(t),\\
\tilde{F}^R_Z(t) &= \sum_n \bigg( k_n \left(q_{n,0} - \alpha(x_0)\right) \cos(\mu_n t)\nonumber\\
& \quad + \mu_n v_{n,0} \sin(\mu_n t)\bigg).
\end{align}
\end{subequations}
The time derivative of $F^R_Z(t,x_t)$ is given by
\begin{align}
\dot{F}^R_Z(t,x_t) &= \tilde{F}^R_Z(t)\frac{\mathrm{d}}{\mathrm{d}t}\alpha^\prime(x_t) + \alpha^\prime(x_t)\frac{\mathrm{d}}{\mathrm{d}t}\tilde{F}^R_Z(t).
\end{align} 
Since the function $\alpha^\prime(x_t)$ depends on time only via $x_t$, its time derivative can be written using the Liouville operator, i.e., $\frac{\mathrm{d}}{\mathrm{d}t}\alpha^\prime(x_t)=L\alpha^\prime(x_t)$. The same is not true for the function $\tilde F^R_Z(t)$. 
By applying the Liouville operator, we find $L \tilde{F}^R_Z(t) = \dot{\tilde{F}}^R_Z(t)-\frac{p_0}{m} \alpha^\prime(x_0)\sum_n k_n \cos(\mu_n t)$. 
Hence, we can write
\begin{subequations}
\begin{align}
\dot{F}^R_Z(t,x_t) &= L\left[ \alpha^\prime(x_t) \tilde{F}^R_Z(t) \right]\nonumber\\
& \quad +\sum_n k_n \alpha^\prime(x_t)\alpha^\prime(x_0)\frac{p_0}{m} \cos(\mu_n t)\\
&= L F^R_Z(t,x_t)\nonumber\\
&\quad + \sum_n k_n \alpha^\prime(x_t)\alpha^\prime(x_0) \dot{x}_0 \cos(\mu_n t).\label{eq_ode_FZ}
\end{align}
\end{subequations}
Eq.~\eqref{eq_ode_FZ} is a linear, inhomogeneous first-order differential equation for $F^R_Z(t,x_t)$. The general solution reads
\begin{subequations}
\begin{align}
F^R_Z(t,x_t) &= e^{tL}F^R_Z(0,x_0)\nonumber\\
&\quad  + \int_0^t\mathrm{d}s\,e^{(t-s)L}\sum_n k_n \alpha^\prime(x_s)\alpha^\prime(x_0)\nonumber\\
& \quad \times \dot{x}_0 \cos(\mu_n(s))\\
%&= e^{tL}F(0) + \int_0^t\mathrm{d}s\,\sum_n k_n \alpha^\prime(x_t)\alpha^\prime(x_{t-s})\dot{x}_{t-s} \cos(\mu_n s)\\
%&= e^{tL}F(0) + \int_0^t\mathrm{d}s\,\sum_n k_n \alpha^\prime(x_t)\alpha^\prime(x_s)\dot{x}_s \cos(\mu_n(t-s))\\
&= e^{tL}F^R_Z(0,x_0)\nonumber\\
& \quad + \int_0^t\mathrm{d}s\,\Gamma\left[t-s,x_t,x_s,\right]\dot{x}_s,\label{eq_sol_FZ}
\end{align}
\end{subequations}
where we used $\Gamma$ from eq.~\eqref{eq_mem_2}. 
By using the Dyson identity from eq.~\eqref{eq_dyson_decomposition} for $e^{tL}$, 
we can write eq.~\eqref{eq_sol_FZ} in terms of the general projection operators  $P$ and $Q$ as
\begin{align}
F^R_Z(t,x_t) &= e^{tQL}F^R_Z(0,x_0) 
+ \int_0^t\mathrm{d}s\,\Gamma\left[t-s, x_t,x_s\right]\dot{x}_s
\nonumber\\
& \quad 
+ \int_0^t\mathrm{d}s\,e^{(t-s)L}PLe^{sQL}F^R_Z(0,x_0).
\label{eq_sol2_FZ}
\end{align}
From eq.~\eqref{eq_AppK_FR}, it follows that
\begin{subequations}
\begin{align}
F^R_Z(0, x_0) &= \alpha^\prime (x_0) \sum_n k_n \left(q_{n,0} - \alpha(x_0)\right)\\
 &= Q_HLp_0 = F^R(0),
\end{align}
\end{subequations}
where we used the definition of the random force  $F^R(t)$ in eq.  \eqref{eq_F_operator}
and the equation of motion for the complementary variables. 
This means that  $F^R_Z(0, x_0)$ coincides with the random force $F^R(t)=e^{tQL}QLp_0$ at time $t=0$. Therefore, by inserting the result in eq.~\eqref{eq_sol2_FZ} into eq.~\eqref{eq_Zwanzig_GLE1}, we obtain eq.~\eqref{eq_Zwanzig_GLE_projection}.
Thus we have proven that the GLE obtained by explicitly solving the non-harmonic Hamiltonian Zwanzig model,
eq.~\eqref{eq_Zwanzig_GLE1}, is equivalent to the GLE obtained from our hybrid projection scheme, eq.~\eqref{eq_Zwanzig_GLE_projection}.

\section{Global Data Smoothing\label{sec_App_smoothing}}

In the main text, we use Legendre polynomial expansions to smooth the data for the dihedral angle dynamics. For $D(A,t)$, the Legendre expansion reads
\begin{align}
\label{eq_polynomial_expansion}
D(A,t) &= \sum_n c_n(t) h_n(A),
\end{align}
where  $h_n$ denotes the  Legendre polynomial of order $n$. The coefficients $c_n(t)$ follow from
\begin{align}
c_n(t) &= \frac{2n+1}{2}\int_{-1}^{1}\mathrm{d}A\,D(A,t)h_n(A).
\end{align}
 In this way, the data is globally smoothed while  spatial symmetries can be conserved. 
 For example, if a function is even  in $A$, only even orders of Legendre polynomials are used
 in the smoothing procedure.

\end{document}